\documentclass[journal]{IEEEtran}
\usepackage{amsmath,graphicx,multirow,booktabs,amssymb,subfigure,cite}

\begin{document}
%
\title{Deep Attention Fusion Feature for Speech Separation with  End-to-End Post-filter Method}
%
%
%

\author{Cunhang~Fan,~\IEEEmembership{Student Member,~IEEE,}
        Jianhua~Tao,~\IEEEmembership{Senior Member,~IEEE,}
        Bin~Liu,~\IEEEmembership{Member,~IEEE,}
        Jiangyan~Yi,~\IEEEmembership{Member,~IEEE,}
        Zhengqi~Wen,~\IEEEmembership{Member,~IEEE,}
        and~Xuefei~Liu,~\IEEEmembership{Member,~IEEE,}
\thanks{This work is supported by the National Key Research \& Development Plan of China (No.2017YFC0820602), the National Natural Science Foundation of China (NSFC) (No.61831022, No.61771472, No.61901473, No.61773379) and Inria-CAS Joint Research Project (No.173211KYSB20170061 and No.173211KYSB20190049).\emph{(Corresponding authors: Jianhua Tao and Bin Liu.)}}
\thanks{The authors are with the National Laboratory of Pattern Recognition, Institute of Automation, Chinese Academy of Sciences, Beijing 100190, China. Cunhang Fan and Jianhua Tao  are also with the School of Artificial Intelligence, University of Chinese Academy of Sciences, Beijing 100190, China. Jianhua Tao is also with the CAS Center for Excellence in Brain Science and Intelligence Technology, Beijing 100190, China. (e-mail:cunhang.fan@nlpr.ia.ac.cn, jhtao@nlpr.ia.ac.cn, liubin@nlpr.ia.ac.cn, jiangyan.yi@nlpr.ia.ac.cn, zqwen@nlpr.ia.ac.cn,xuefei.liu@nlpr.ia.ac.cn)}
}

%
%

\markboth{Journal of \LaTeX\ Class Files,~Vol.~14, No.~8, August~2015}
{Shell \MakeLowercase{\textit{et al.}}: Bare Demo of IEEEtran.cls for IEEE Journals}

%



\maketitle

\begin{abstract}
In this paper, we propose an end-to-end post-filter method with deep attention fusion features for monaural speaker-independent speech separation. At first, a time-frequency domain speech separation method is applied as the pre-separation stage. The aim of pre-separation stage is to separate the mixture preliminarily. Although this stage can separate the mixture, it still contains the residual interference. In order to enhance the pre-separated speech and improve the separation performance further, the end-to-end post-filter (E2EPF) with deep attention fusion features is proposed. The E2EPF can make full use of the prior knowledge of the pre-separated speech, which contributes to speech separation. It is a fully convolutional speech separation network and uses the waveform as the input features. Firstly, the 1-D convolutional layer is utilized to extract the deep representation features for the mixture and pre-separated signals in the time domain. Secondly, to pay more attention to the outputs of the pre-separation stage, an attention module is applied to acquire deep attention fusion features, which are extracted by computing the similarity between the mixture and the pre-separated speech. These deep attention fusion features are conducive to reduce the interference and enhance the pre-separated speech. Finally, these features are sent to the post-filter to estimate each target signals. Experimental results on the WSJ0-2mix dataset show that the proposed method outperforms the state-of-the-art speech separation method. Compared with the pre-separation method, our proposed method can acquire 64.1\%, 60.2\%, 25.6\% and 7.5\% relative improvements in scale-invariant source-to-noise ratio (SI-SNR), the signal-to-distortion ratio (SDR), the perceptual evaluation of speech quality (PESQ) and the short-time objective intelligibility (STOI) measures, respectively.

\end{abstract}


%
\IEEEpeerreviewmaketitle

\section{Introduction}
%
%
%
%

\IEEEPARstart{S}{peech} separation aims to estimate the target sources from a noisy mixture, which is known as the cocktail party problem \cite{O2015Attentional,fan2020spatial}. As for monaural speech separation, it is a very challenging task because only single channel can be used. This study focuses on monaural speaker-independent speech separation. 

Recently, deep learning has been applied to address speaker-independent speech separation, which has obtained impressive results \cite{wang2018supervised,wang2019deep,liu2018casa,wang2018deep,luo2018tasnet,xu2018single,fan2018Utterance,wang2019apitch}. The difficulty of speaker-independent speech separation is label ambiguity or permutation problem \cite{Kolbaek2017Multitalker, Hershey2016Deep}. In order to deal with this problem, deep clustering (DC) \cite{Hershey2016Deep} is proposed, which is a state-of-the-art method for speaker-independent speech separation. DC is usually formulated as two-step processes: embedding learning and embedding clustering. Firstly, as for embedding learning, a bidirectional long-short term memory (BLSTM) network is trained to project each time-frequency (T-F) bin of mixture spectrogram into an embedding vector. The training objective is the Frobenius norm between the affinity matrices of the embedding vector and the ideal binary mask. In this way, if the T-F bins belong to the same speaker, these embedding vectors are grouped closer together. Otherwise, they become farther apart. Finally, in order to acquire the binary mask of each source, K-means algorithm is applied to cluster these embedding vectors, which is the embedding clustering. Although DC gets good performance, it still has two limitations. Firstly, the training objective is defined in the embedding vectors, instead of the real separated sources. These embedding vectors do not necessarily imply perfect separation of the sources in the signal space. Secondly, DC applies the unsupervised K-means clustering algorithm to estimate the binary masks of target sources. Therefore, the performance of speech separation is limited by the K-means clustering algorithm. To overcome the training objective limitation of DC, the deep attractor network (DANet) \cite{chen2017deep} method is proposed. Same as DC, the DANet also maps the mixture spectrogram into a high-dimensional embedding space. Different from DC, DANet firstly creates attractor points at the embedding space. Then the similarities between the embedded points and each attractor are applied to estimate each source's mask. However, at the test stage, it still requires the unsupervised K-means clustering algorithm to acquire the binary mask.

Frame-level permutation invariant training (PIT) \cite{Yu2017Permutation} deals with the permutation problem in a different way. During training, the frame-level PIT (denoted by tPIT) computes all possible label permutations for each frame. Then tPIT uses the permutation with the lowest mean square error (MSE) as the loss to train the separation model. It can get a good performance for frame-level separation. However, in the real-world conditions, the frame-level permutation of separated signals is unknown. It means that tPIT needs the speaker tracing step during inference. To address this issue, utterance-level PIT (uPIT) \cite{Kolbaek2017Multitalker} is proposed. With uPIT, instead of choosing the permutation at frame-level, the permutation corresponding to the minimum utterance-level separation error is used for all frames in one utterance. In this way, uPIT can effectively eliminate the speaker tracing problem. However, tPIT and uPIT only reduce the distance between the same speakers, they don't increase the distance between the different speakers. This may lead to increasing the possibility of remixing the separated sources.

In order to use both of DC and PIT, Chimera++ network \cite{wang2018alternative} is applied for speech separation, which is followed by the Chimera network \cite{luo2017deep}. The Chimera++ network uses a multi-task learning architecture to combine the DC and PIT. However, it simply employs the DC and PIT as two outputs of the separation model rather than fuses them deeply. Therefore, it does not solve the limitations of DC and PIT. Computational auditory scene analysis (CASA) \cite{rouat2008computational} is a traditional speech separation method, which is inspired by human auditory scene analysis. Deep CASA \cite{liu2018casa} is another method to combine the DC and PIT. It adopts the same divide-and-conquer strategy of CASA. Deep CASA is a two-stage speech separation method. Firstly, tPIT is used to estimate each source from the mixture spectrogram. Then, DC is used as the speaker tracing step. In other words, DC is applied to estimate the optimized permutation at frame-level. Although deep CASA acquires good separation performance, it is also limited by the K-means algorithm.

Motivated by PIT, DC and discriminative learning \cite{fan2018Utterance,Grais2016Combining,huang2014singing,Huang2014Deep,fan2020spatial}, we proposed a discriminative learning method for speaker-independent speech separation with deep embedding features (denoted by uPIT+DEF+DL) in our previous work \cite{fan2019discriminative}. uPIT+DEF+DL combines DC and PIT in a deep fusion method and addresses the limitations of DC and PIT very well. It utilizes the DC network as the extractor of deep embedding features. Then instead of using K-means clustering algorithm to estimate the target sources, uPIT+DEF+DL applies the uPIT to separate the speech from these deep embedding features. Although uPIT+DEF+DL can separate the mixture well, it still has two drawbacks limiting its performance. Firstly, it uses the separated magnitude and mixture phase to reconstruct target signals by inverse short-time Fourier transformation (ISTFT), which is mismatched for magnitude and phase. Secondly, the separated signals by the uPIT+DEF+DL may still contain the residual interference signals, which damages the performance of speech separation.

In this study, in order to address the above issues, we propose an end-to-end post-filter (E2EPF) method with deep attention fusion features for monaural speaker-independent speech separation. The proposed E2EPF utilizes the time-domain waveform as the input features. The waveform contains all of the information of the raw wave, including the magnitude and phase. Therefore, separating the speech from waveform can solve the mismatch problem of magnitude and phase. At the first, the uPIT+DEF+DL is used as the pre-separation stage to preliminarily estimate target sources from the mixture spectrogram through T-F domain. The separated speech by this stage may still contain the residual interference. To further enhance the pre-separated speech, the E2EPF with deep attention fusion features is applied. The E2EPF can make full use of the prior knowledge of pre-separated speech to help reduce the residual interference. Firstly, the mixture and pre-separated signals are processed by the 1-D convolutional layer to extract deep representation features. Secondly, instead of simply stacking these deep representation features, an attention module is applied to compute the similarity between the mixture and the pre-separated speech, which is used as the extractor of deep attention fusion features. These features can make the proposed model pay more attention to the pre-separated signals so that the proposed E2EPF can reduce the interference more easily and enhance the pre-separated speech. 


The main contribution of this paper is two-fold. Firstly, we propose the E2EPF to further enhance the pre-separated speech and reduce the residual interference. Secondly, deep attention fusion features are applied to compute the similarity between the mixture and the pre-separated speech. Experiments are conducted on WSJ0-2mix and WSJ0-3mix datasets \cite{Hershey2016Deep}. Experimental results show that our proposed method outperforms the state-of-the-art speech separation method.

The rest of this paper is organized as follows. Section \(\rm \uppercase\expandafter{\romannumeral2}\) presents discriminative learning for monaural speech separation using deep embedding features. Section \(\rm \uppercase\expandafter{\romannumeral3}\) introduces the proposed end-to-end post-filter speech separation method. The experimental setup is stated in section \(\rm \uppercase\expandafter{\romannumeral4}\). Section \(\rm \uppercase\expandafter{\romannumeral5}\) shows experimental results. Section \(\rm \uppercase\expandafter{\romannumeral6}\) shows the discussions. Section \(\rm \uppercase\expandafter{\romannumeral7}\) draws conclusions.

%

\section{Discriminative Learning for Monaural Speech Separation Using Deep Embedding Features}
\label{sec:separation}

The object of monaural speech separation is to estimate target sources from the mixture speech recorded by single channel.
\begin{equation}
\textbf{y}(t) = \sum_{s=1}^{S}{\textbf{x}_s(t)}
\label{eq1}
\end{equation}
where \(\textbf{y}(t)\) is the mixture speech, \(t\) is the time index, \(S\) is the number of sources and \(\textbf{x}_s(t)\), \(s=1,...,S\) are target sources. And the corresponding short-time Fourier transformation (STFT) of \(\textbf{y}(t)\) and \(\textbf{x}_s(t)\) are \(Y(t,f)\) and \(X_s(t,f)\).

The speech separation aims to estimate each source signals \(\textbf{x}_s(t)\) from \(\textbf{y}(t)\) or \(Y(t,f)\). In this section, we introduce the discriminative learning method for speech separation with deep embedding features \cite{fan2019discriminative}, which is based on uPIT. This method is denoted as uPIT+DEF+DL. We use this method as our pre-separation stage and our baseline.

\subsection{Deep Embedding Features}

\begin{figure}[t]
	\centering
	\includegraphics[width=\linewidth]{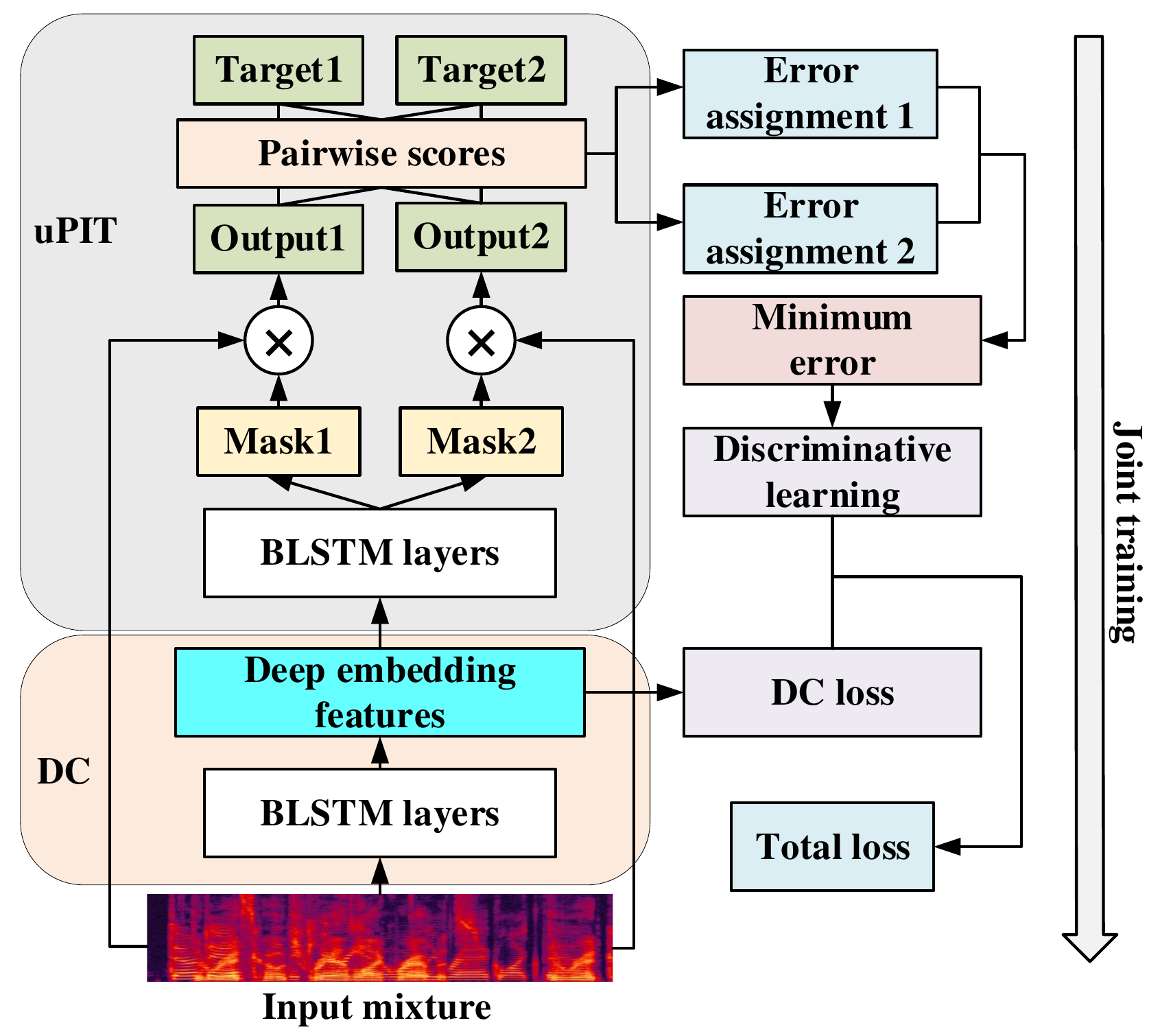}
	\caption{Schematic diagram of uPIT+DEF+DL speech separation system. DC loss is the loss of deep clustering.}
	\label{fig:DC_uPIT_system}
\end{figure}

Fig.~\ref{fig:DC_uPIT_system} shows the schematic diagram of uPIT+DEF+DL speech separation system. Firstly, a BLSTM network is trained as the extractor of deep embedding features (DEF). The aim of the extractor is to project the mixed amplitude spectrum \(|Y(t,f)|\) of each T-F bin into the D-dimensional deep embedding features \(V\).
\begin{equation}
V=\gamma_{\theta}(|Y(t,f)|) \in{\mathbb{R}^{TF\times{D}}}
\label{eq7}
\end{equation}
where TF is the number of T-F bins and \(\gamma_{\theta}(*)\) is the BLSTM mapping function. Here we consider a unit-norm embedding, so
\begin{equation}
|\textbf{v}_i|^2=1,\quad \textbf{v}_i={\textbf{v}_{i,d}}
\label{eq8}
\end{equation}
where \(\textbf{v}_{i,d}\) is the value of the \(d\)-th dimension of the embedding for element \(i\). We let the embeddings \(V\) to implicitly represent an \(TF\times{TF}\) estimated affinity matrix \(VV^T\).

As for the deep embedding features extractor, the loss function \(J_{DC}\) is defined as follow:
\begin{equation}
\begin{split}
J_{DC} &=||VV^T-BB^T||_F^2\\
&=||VV^T||_F^2-2||V^TB||_F^2+||BB^T||_F^2\\
\end{split}
\label{eq9}
\end{equation}
where \(B\in{\mathbb{R}^{TF\times{S}}}\) is a binary matrix, which means the source membership function for each T-F bin. Specifically, if the energy of source \(s\) is the highest compared with other sources, \(B_{tf,s}=1\). Otherwise, \(B_{tf,s}=0\). S denotes the source number. \(||*||_F^2\) is the squared Frobenius norm. 

\subsection{uPIT Based Speech Separation Model with Deep Embedding Features}

As for DC \cite{Hershey2016Deep}, the training objective is not the real separated sources. Besides, the unsupervised K-means clustering algorithm is applied to acquire binary masks. Therefore, the performance is limited by the K-means algorithm. In order to address these issues, we use the deep embedding vectors extracted by DC as the input of uPIT to directly learn each source's soft masks. In this way, on one hand, we directly use the real separated sources as the training objective. In other words, the DC and uPIT can be trained end-to-end. On the other hand, the performance of speech separation is not limited by the K-means algorithm.


Phase sensitive mask (PSM) \cite{Wang2014On, Erdogan2015Phase} is proved to be effective for speech separation because it makes full use of the phase information \cite{Kolbaek2017Multitalker}. In this paper, we utilize the PSM for speech separation in the T-F domain. The ideal PSM is defined as:
\begin{equation}
M_s(t,f)=\frac{|X_s(t,f)|cos(\theta_y(t,f)-\theta_s(t,f))}{|Y(t,f)|}
\label{eq4}
\end{equation}
where \(\theta_y(t,f)\) and \(\theta_s(t,f)\) are the phase of mixture speech and target source \(s\).

uPIT computes the MSE for all possible speaker permutations at utterance-level. Then the minimum cost among all permutations (P) is chosen as the optimal assignment.
\begin{equation}
J_{uPIT}=\mathop{arg\,min}_{\theta_s\in{P}}\sum_{s=1}^S|||Y|\odot{\widetilde{M}_s}-|X_{\theta_s}|cos(\theta_y-\theta_s)||_F^2
\label{eq6}
\end{equation}
where the number of all permutations (P) is \(N=S!\) (\(!\) denotes the factorial symbol). The \((t,f)\) is omitted in \(\widetilde{M}_s\), \(Y\), \(X\), \(\theta_y\) and \(\theta_s\).

\subsection{Discriminative Learning}

For uPIT, the target of minimizing Eq.\ref{eq6} is to reduce the distance between the outputs and their corresponding target sources. To decrease the possibility of remixing separated sources, the discriminative learning (DL) is applied to our proposed model. DL not only reduces the distance between the prediction and the corresponding target, but also increases the distance between the prediction and the interference sources.

We assume that \(\phi^*\) is the chosen permutation (the same as the \(J_{uPIT}\) in Eq.~\ref{eq6}), which has the lowest MSE among all permutations. Therefore, the discriminative learning loss function can be defined as:
%
\begin{equation}
J_{DL}=\phi^*-\sum_{\phi \ne{\phi^*},\phi \in{P}}\alpha \phi
\label{eq11}
\end{equation}
where \(\phi\) is a permutation from \(P\) but does not contain \(\phi^*\), \(\alpha \ge0\) is the regularization parameter of \(\phi\). When \(\alpha=0\), the loss function is the same as the \(J_{uPIT}\) in Eq.~\ref{eq6}. It means with no discriminative learning.




\subsection{Joint Training}

To extract embedding features effectively, we apply the joint training framework to the proposed system. The loss function of joint training is defined as follow:
\begin{equation}
\begin{split}
J & =\lambda{J_{DC}}+(1-\lambda)J_{DL}\\
& = \lambda{J_{DC}}+(1-\lambda)(\phi^*-\sum_{\phi \ne{\phi^*},\phi \in{P}}\alpha \phi)
\end{split}
\label{eq13}
\end{equation}
where \(\lambda\in{[0,1]}\) controls the weight of \(J_{DC}\) and \(J_{DL}\). 


\section{the proposed Speech Separation method}
\label{sec:proposed method}

\begin{figure*}
	\centering
	\subfigure[End-to-end post-filter diagram]{
		\includegraphics[width=0.8\textwidth]{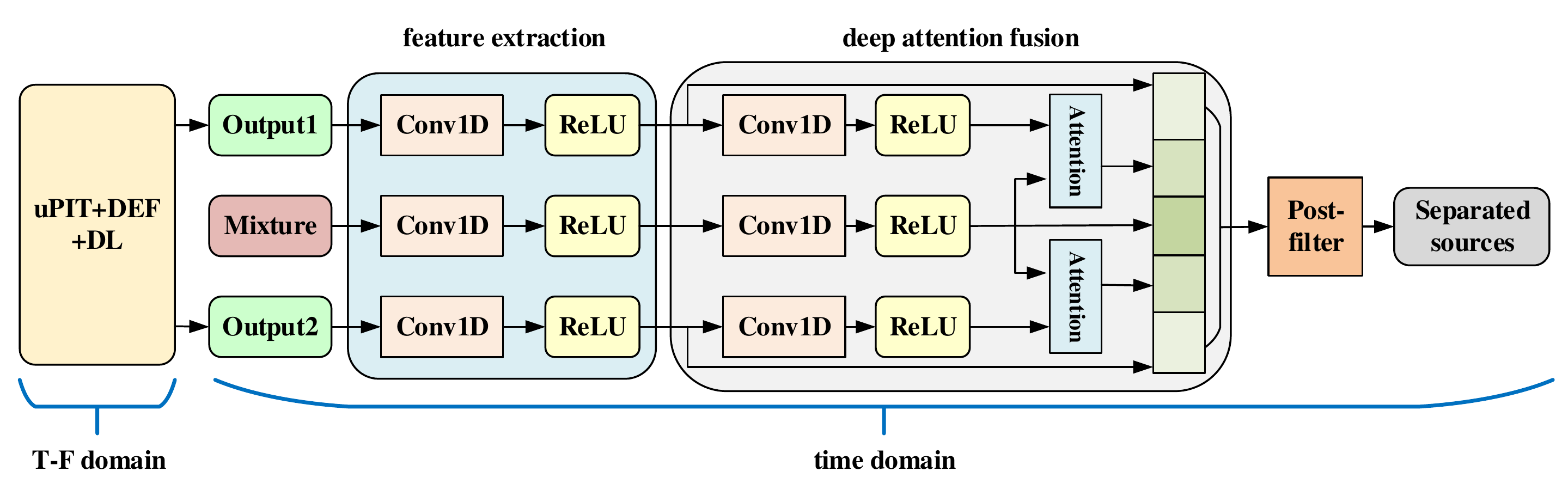}
	}
	
	\centering
	\subfigure[Post-filter]{
		\includegraphics[width=0.68\textwidth]{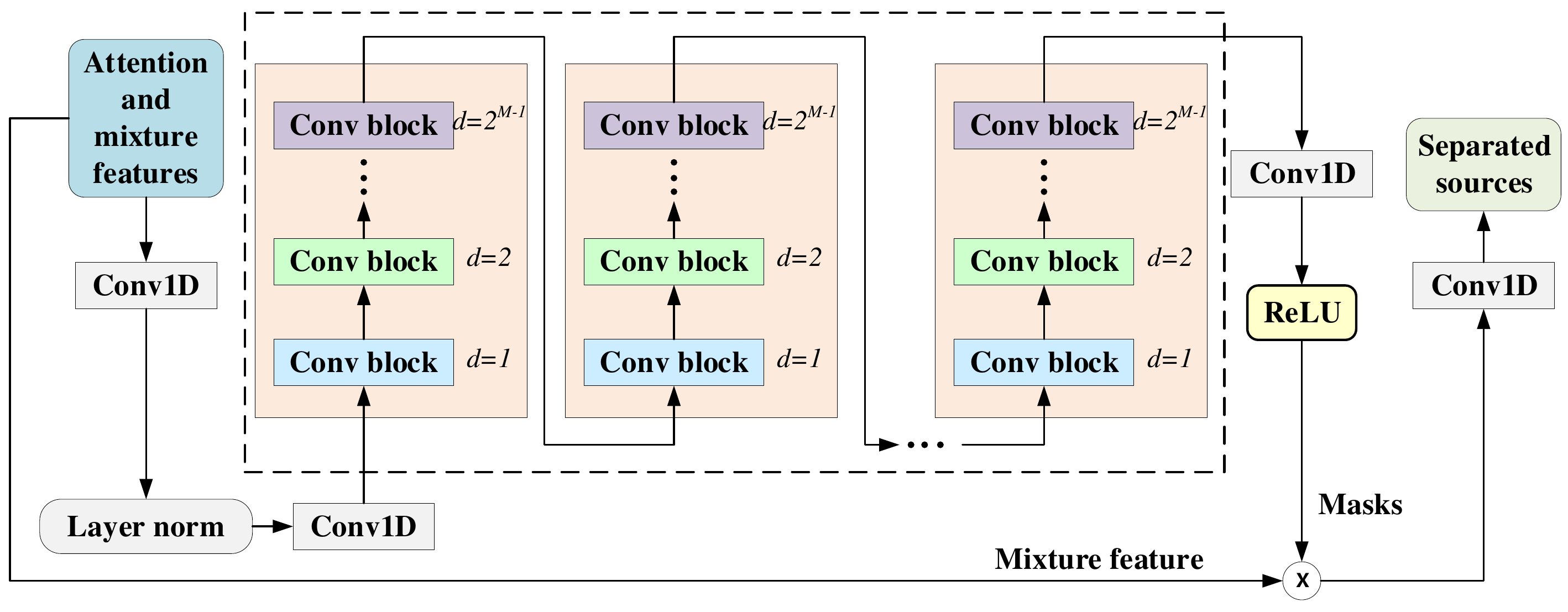}
	}
	\subfigure[Conv block]{
		\begin{minipage}[t]{0.29\textwidth}
			\centering
			\includegraphics[width=0.58\textwidth]{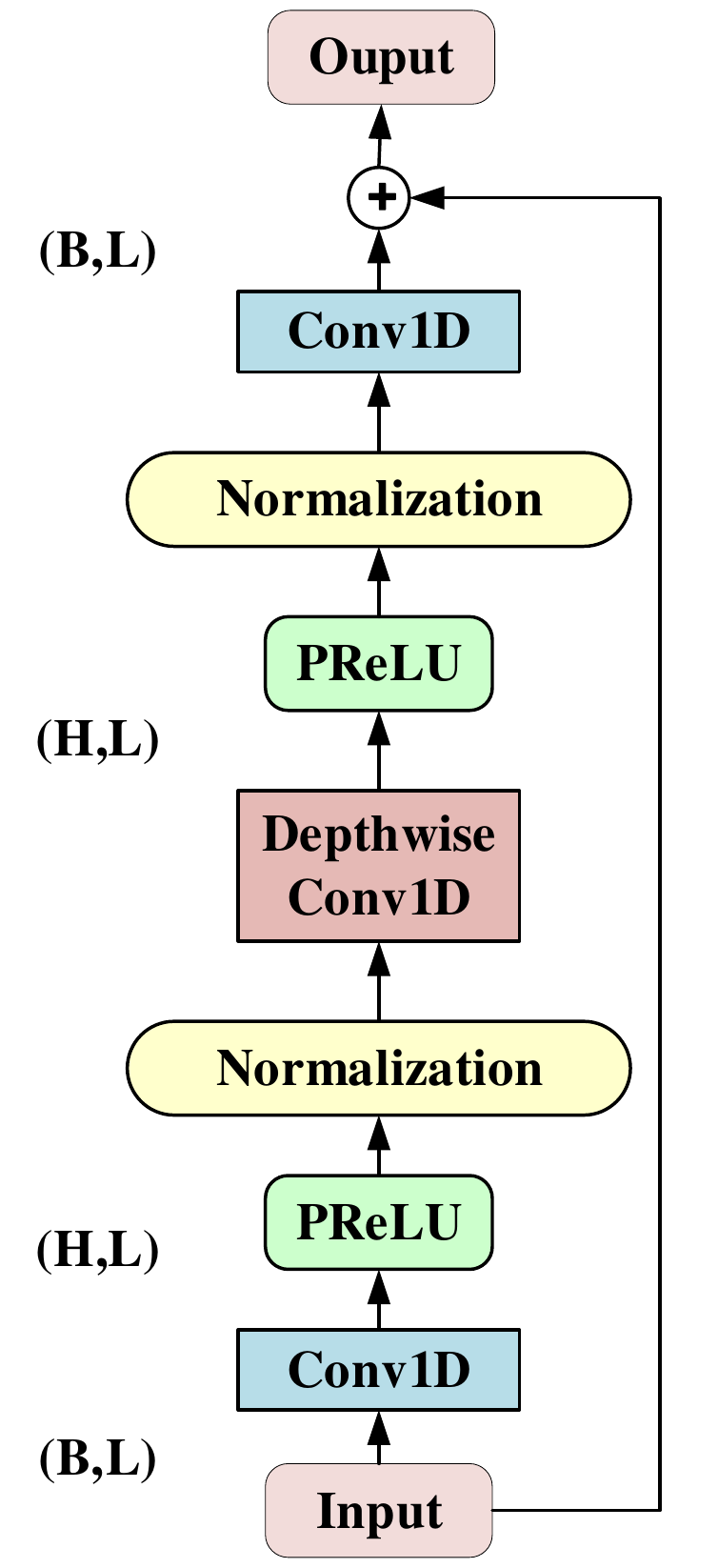}
	\end{minipage}}
	\caption{(a): the diagram of the end-to-end post-filter. It contains three parts: feature extraction, deep attention fusion and post-filter. Features are extracted by the 1-D convolution operation. Then attention mechanism is leveraged to the deep attention fusion. Finally, these features are inputted to the post-filter for speech separation. (b): the detail block diagram of post-filter. The post-filter is composed of 1-D convolution and temporal convolutional network (TCN). (c): the design of 1-D convolution block.}
	\label{fig:End-to-end Post-filter Stage} 
\end{figure*}

%
%
%

In this paper, we propose an end-to-end post-filter (E2EPF) with deep attention fusion features for monaural speaker-independent speech separation. Firstly, we use the uPIT+DEF+DL to separate the mixture preliminarily in the T-F domain, which is used as the pre-separation stage. The separated speech by this method may still contain the residual interference. In order to further enhance the separated speech and improve the performance of speech separation, we utilize the E2EPF with deep attention fusion features as another stage. The E2EPF can make full use of the prior knowledge of the pre-separated speech. The E2EPF is a fully convolutional network and applies the waveform as the input feature. Besides, in order to make the separation model pay more attention to the pre-separated signals, an attention module is utilized to extract deep attention fusion features, which are computed the similarity between the mixture and pre-separated signals. 

The E2EPF mainly solves two problems. Firstly, in the pre-separation stage, it only enhances the magnitude and leaves the phase spectrum unchanged. The mismatched magnitude and phase are used to reconstruct estimated signals, which damages the performance of speech separation. The E2EPF does the speech separation in the time domain so that it can enhance the magnitude and phase spectrum simultaneously. Secondly, the separated signals by the pre-separation stage may still contain the residual interference. The E2EPF makes full use of the prior knowledge of the pre-separated speech and applies the deep attention fusion features to further remove the residual interference and improve the performance of speech separation.

The E2EPF utilizes the waveform as the input features. It consists three parts: feature extraction, deep attention fusion and post-filter, as shown in Fig.~\ref{fig:End-to-end Post-filter Stage} (a). This section we will introduce these three parts detailedly.

\subsection{Feature Extraction}

The input mixture speech (\(\textbf{y}(t)\)) and the output sources (\(\textbf{o}_s(t)\), \(s=1,2,...,S\)) of the pre-separation stage can be divided into overlapping segments of length \(L\). We denote them as \(\textbf{y}_k\in\mathbb{R}^{1\times L} \) and \(\textbf{o}_{sk}\in\mathbb{R}^{1\times L} \), where \(k=1,...,\hat{T}\) is the index of segment and \(\hat{T}\) denotes the total number of segments in \(\textbf{y}(t)\) and \(\textbf{o}_s(t)\). 

The 1-D convolution operation is used to extract deep features from the \(\textbf{y}\) and \(\textbf{o}_{s}\) (we drop the index \(k\) and time \(t\) from now on).
\begin{equation}
\textbf{w}_y=ReLU(\textbf{y}{U}_y)
\label{eq15}
\end{equation}
\begin{equation}
\textbf{w}_s=ReLU(\textbf{o}_s{U}_s),\  s=1,2,...,S
\label{eq16}
\end{equation}
where \(\textbf{w}_y, \textbf{w}_s\in\mathbb{R}^{1\times N} \) are the deep features extracted from the \(\textbf{y}\) and \(\textbf{o}_{s}\), respectively. \({U}_y\in\mathbb{R}^{N\times L} \) and \({U}_s\in\mathbb{R}^{N\times L} \) are the basis functions of 1-D convolution operation, which contains \(N\) vectors with length \(L\) each. \(ReLU(*)\) denotes the rectified linear unit, which is an optional nonlinear function.

\subsection{Deep Attention Fusion}


Recently, attention models have been successfully applied to the sequence-to-sequence learning tasks \cite{bahdanau2014neural,bahdanau2016end,hao2019attention,luong2015effective,xiao2019single}. In this study, attention mechanism is leveraged to acquire the deep attention fusion features. 

The aim of the attention mechanism is to make the separation model pay more attention to the output signals of the pre-separation stage. It is used to compute the similarity between the mixture and pre-separated signals. Therefore, the E2EPF can further reduce the interference signals and improve the performance of speech separation. In order to compute the similarity between mixture and the pre-separated signals, the \(\textbf{w}_y\) and \(\textbf{w}_s\) are sent to another 1-D convolutional layer.
\begin{equation}
\textbf{w}_y^{'}=ReLU(\textbf{w}_y{U}_y^{'})
\label{eq155}
\end{equation}
\begin{equation}
\textbf{w}_s^{'}=ReLU(\textbf{w}_s{U}_s^{'}),\  s=1,2,...,S
\label{eq166}
\end{equation}
where \({U}_y^{'}\in\mathbb{R}^{N\times L} \) and \({U}_s^{'}\in\mathbb{R}^{N\times L} \) are the basis functions of 1-D convolution operation.

According to the global attention mechanism \cite{luong2015effective}, the attention weight \(\alpha_{t,t^{'}}\) can be learned:
\begin{equation}
\alpha_{t,t^{'}}=\frac{{\rm exp}(d_{t,t^{'}})}{\sum_t^{'}{{\rm exp}(d_{t,t^{'}})}}
\label{eq17}
\end{equation}
where \(d_{t,t^{'}}\) is the correlation between \(\textbf{w}_y^{'}\) and \(\textbf{w}_s^{'}\), which measures their similarity. The attention weight \(\alpha_{t,t^{'}}\) is the softmax of \(d_{t,t^{'}}\) over \(t^{'}\in{[1,N]} \) . We follow the dot-based function in \cite{luong2015effective} as the \(d_{t,t^{'}}\). \(d_{t,t^{'}}\) is defined as follow:
\begin{equation}
d_{t,t^{'}}=\textbf{w}_y^{'T}\textbf{w}_s^{'}
\label{eq18}
\end{equation}

The context vector \(\textbf{c}_{t^{'}s}\in\mathbb{R}^{1\times N}\) can be calculated by the weighted average of \(\textbf{w}_s^{'}\):
\begin{equation}
\textbf{c}_{t^{'}s}=\sum_t{\alpha_{t,t^{'}}\textbf{w}_s^{'}}
\label{eq19}
\end{equation}

As shown in Fig.~\ref{fig:End-to-end Post-filter Stage} (a), gray area is the deep attention fusion part. Finally, these two context vectors \(\textbf{c}_{t^{'}s}\)  and the mixture deep feature \(\textbf{w}_y^{'}\) are as the deep attention fusion features to the next post-filter part.

\subsection{Post-filter}

The detail block diagram of post-filter is shown in Fig.~\ref{fig:End-to-end Post-filter Stage} (b), which adopts the temporal convolutional network (TCN) similar to TasNet \cite{luo2018arXivtasnet}. TCN is leveraged to the end-to-end post-filter, which has shown comparable even better performance than RNNs in various sequence modeling tasks \cite{bai2018empirical,Lea2016Temporal,pandey2019tcnn,liu2019divide,luo2019conv,luo2018arXivtasnet,lea2017temporal}. The post-filter is a fully-convolutional module including stacked dilated 1-D convolutional blocks as shown in Fig.~\ref{fig:End-to-end Post-filter Stage} (c). Compared with the TasNet \cite{luo2018arXivtasnet}, there are two main differences. Firstly, our proposed post-filter makes full use of the prior knowledge of the pre-separated speech and the post-filter is used as the second stage to improve the separation performance. Secondly, to pay more attention to the pre-separated speech, these deep attention fusion features are applied.

TCNs are used to replace for recurrent neural networks (RNNs), which have shown comparable even better performance than RNNs in various sequence modeling tasks \cite{bai2018empirical,Lea2016Temporal,pandey2019tcnn,liu2019divide,luo2019conv,luo2018arXivtasnet}. For each TCN, 1-D convolutional blocks have increasing dilation factors (\(1,2,...,2^{M-1}\), \(M\) is the number of convolutional blocks), as shown in the light brown of Fig.~\ref{fig:End-to-end Post-filter Stage} (b). These increasing dilation factors can capture a large temporal context. To further increase the receptive field, the \(M\) stacked dilated convolutional blocks are repeated \(R=4\) times.

Fig.~\ref{fig:End-to-end Post-filter Stage} (c) shows the stacked dilated 1-D convolutional block, which follows \cite{vanwavenet}. To avoid losing input information, the skip connection is utilized between the input and the next block. The depthwise separable convolution has been proven to be effective for image processing tasks \cite{chollet2017xception,howard2017mobilenets}. Then, the depthwise separable convolution is applied to further decrease the parameters numbers. A nonlinear activation function and a normalization operation are added after both the first \(1\times{1}-conv\) and \(D-conv\) blocks respectively. The parametric rectified linear unit (PReLU) \cite{he2015delving} is applied. The reason is that PReLU can improve model fitting with nearly zero extra computational cost and little overfitting risk \cite{he2015delving}. The type of the normalization is the global layer normalization (gLN) because that the gLN outperforms all other normalization methods \cite{luo2019conv}.

The output of the stacked dilated 1-D convolutional block is inputted to a 1-D convolutional layer with ReLU nonlinear function and we denote these neural networks as \(\mathcal{\gamma}(*)\). The reason of using ReLU is that we want the network to learn target masks like the T-F domain. The output of \(\mathcal{\gamma}(*)\) is the estimated mask \(\textbf{m}_s\in\mathbb{R}^{1\times N}\) of each source similar to the pre-separation stage.
\begin{equation}
\textbf{m}_s=\mathcal{\gamma}([\textbf{w}_s,\textbf{c}_{t^{'}s};\textbf{w}_y^{'}]),\  s=1,2,...,S
\label{eq20}
\end{equation}

Then the separated representation \(\textbf{e}_s\) of source \(s\) can be estimated as following:
\begin{equation}
\textbf{e}_s=\textbf{w}_y\odot{\textbf{m}_s}
\label{eq21}
\end{equation}
where \(\odot\) denotes the element-wise multiplication.

Finally, the estimated waveform of source \(s\) \(\widetilde{\textbf{x}}_s\) is reconstructed by the transposed 1-D convolution operator:
\begin{equation}
\widetilde{\textbf{x}}_s=\textbf{e}_sU_e
\label{eq22}
\end{equation}
where \(U_e\in\mathbb{R}^{N\times L}\) denotes the basis function of transposed 1-D convolution operator.

\subsection{Training Objective}

In order to improve the separation performance, the training objective of the end-to-end post-filter is to maximize the scale-invariant source-to-noise ratio (SI-SNR) \cite{vincent2006performance}. The SI-SNR is defined as:
\begin{equation}
\textbf{x}_{target}=\frac{\langle\widetilde{\textbf{x}},\textbf{x}\rangle\textbf{x}}{\|\textbf{x}\|^2}
\label{eq23}
\end{equation}
\begin{equation}
\textbf{e}_{noise}=\widetilde{\textbf{x}}-\textbf{x}_{target}
\label{eq24}
\end{equation}
\begin{equation}
\textbf{\rm SI-SNR}=10log_{10}\frac{\|\textbf{x}_{target}\|^2}{\|\textbf{e}_{noise}\|^2}
\label{eq25}
\end{equation}
where \(\widetilde{\textbf{x}}\) and \(\textbf{x}\) denote the estimated and target sources, respectively. \(\|\textbf{x}\|^2=\langle\textbf{x},\textbf{x}\rangle \) is the signal power. In order to solve the permutation problem, the uPIT is utilized during training.

\section{Experimental Setup}

\subsection{Dataset}

The WSJ0-2mix and WSJ0-3mix datasets \cite{Hershey2016Deep} are used to conduct our experiments, which is derived from WSJ0 corpus \cite{garofalo2007csr}. It has training, validation and test set. The training set has 20,000 utterances about 30 hours. It is 5,000 utterances about 10 hours for validation set. As for test set, it has 3,000 utterances about 5 hours. All of the data is generated by randomly selecting utterances from WSJ0 set with signal-to-noise ratios (SNRs) between -5dB and 5dB. The training and validation set are generated from the WSJ0 training set (\texttt{si\_tr\_s}). The test set is generated from the WSJ0 development set (\texttt{si\_dt\_05}) and evaluation set (\texttt{si\_et\_05}). All the waveforms are sampled at 8000 Hz.


In order to evaluate the separation performance, the validation set is used as the closed conditions (CC) and the test set is used as the open condition (OC). 


\subsection{Baseline model}

In this paper, we use the uPIT+DEF+DL as our baseline model. To compute the short-time Fourier transform (STFT), the hamming window is 32ms and window shift is 16ms. Therefore, the dimension of the spectral magnitude is 129. We use the normalized amplitude spectrum of the mixture speech as the input features. 


There are two BLSTM layers with 896 units as for the extractor of deep embedding features. We set the dimension of embedding D to 40. Following the embedding layer, a tanh activation function is utilized. For uPIT separation network, there is only one layer with 896 units. Therefore, the network of pre-separation has 3 BLSTM layers in total, which is the same as the baseline in \cite{Kolbaek2017Multitalker}. As for the mask estimation layer, a Rectified Liner Uint (ReLU) activation function is used to estimate the mask of each source, which is followed by the uPIT separation network. The discriminative learning parameter \(\alpha\) is set to 0.1.


For each BLSTM layer, a random dropout is applied and dropout rate is set to 0.5. The batch-size is 16 utterances which is generated by randomly selecting. The minimum epoch is set to 30. The learning rate is initialized as 0.0005. When the training loss increases on the validation set, the learning rate is scaled down by 0.7. When the relative loss improvement is lower than 0.01, the model is early stopped. The models of this stage are optimized with the Adam algorithm \cite{Kingma2014Adam}.

In this paper, we re-implement uPIT \cite{Kolbaek2017Multitalker} with our experimental setup, which has three BLSTM layers with 896 units. The others are the same as the experimental setup of our pre-separation stage.


\begin{table*}[t]
	\caption{The results of SDR, SIR, SAR and PESQ for different separation methods with closed (CC) and open (OC) condition on WSJ0-2mix dataset. \(\lambda\) is the weight of joint training in Eq.\ref{eq13}. DEF denotes the deep embedding features. uPIT is the baseline method, uPIT+DEF and uPIT+DEF+DL are our proposed methods. uPIT+DEF means with no discriminative learning.}
	\label{tab:results1}
	\centering
	\begin{tabular}{c |c|cc|cc|cc|cc|cc|cc|cc|cc}
		\toprule
		\multicolumn{1}{c}{\multirow{3}*{Method}} & \multicolumn{1}{|c|}{\multirow{3}*{\(\lambda\)}} & \multicolumn{8}{|c|}{Optimal (Opt.) Assign.} & \multicolumn{8}{|c}{Default (Def.) Assign.} \\
		\cline{3-18}
		& & \multicolumn{2}{|c|}{SDR(dB)} & \multicolumn{2}{|c|}{SIR(dB)} & \multicolumn{2}{c}{SAR(dB)} & \multicolumn{2}{|c|}{PESQ} & \multicolumn{2}{|c|}{SDR(dB)} & \multicolumn{2}{|c|}{SIR(dB)} & \multicolumn{2}{c}{SAR(dB)} & \multicolumn{2}{|c}{PESQ}\\
		\cline{3-18}
		& & CC& OC& CC& OC& CC& OC & CC& OC& CC& OC& CC& OC& CC& OC& CC& OC\\
		\midrule
		uPIT&     -&   11.3& 11.2&  18.8& 18.8& 12.3& 12.3&  2.68&   2.67&  10.3 & 10.1& 17.7 & 17.5& 11.5 & 11.3& 2.60& 2.58\\
		\midrule
		uPIT+DEF&  0.01& 11.7& 11.6& 19.4& 19.5& 12.7& 12.6& \textbf{2.85}& \textbf{2.84}& 10.8 & 10.7&  18.4& 18.4 & \textbf{12.0}&  11.8& \textbf{2.77}&\textbf{2.75}\\
		uPIT+DEF&  0.05& 11.7& 11.7& 19.5& 19.6& 12.7& 12.6& {2.84}& \textbf{2.84}&   10.8& \textbf{10.8}& 18.4& \textbf{18.8}& 11.9& \textbf{11.9} & {2.76}&\textbf{2.75}\\
		uPIT+DEF&  0.1& 11.7& 11.7& 19.5& 19.5& 12.7& 12.6& {2.84}& \textbf{2.84}&   10.8& 10.7& 18.5& 18.4& \textbf{12.0}& \textbf{11.9} & 2.76&2.74\\
		\midrule
		uPIT+DEF+DL&  0.05& \textbf{11.9}& \textbf{11.9}& \textbf{19.9}& \textbf{20.0}& \textbf{12.8}& \textbf{12.7}& 2.83& 2.83&   \textbf{11.0}& \textbf{10.8}& \textbf{18.8}& \textbf{18.8}& \textbf{12.0}& \textbf{11.9} & 2.74 &2.73\\
		\bottomrule
	\end{tabular}
\end{table*}

\subsection{The proposed end-to-end post-filter method}

The length of input waveform is 4-second long segments. The learning rate is initialized as 0.0001. If the training loss increases in 3 consecutive epochs on the validation set, the learning rate is halved. Same as the pre-separation stage, the optimizer of this stage is the Adam algorithm \cite{Kingma2014Adam}. The maximum number of epoch is 100. As for the feature extraction, the number of the first 1-D convolutional filters is 256 with length 20 (in samples) (\(N=256, L=20\) in Section \ref{sec:proposed method}-B). As for the other 1-D convolution, the number of channels all is 256. For convolutional blocks, The numbers of channels and kernel size are 512 and 3, respectively. The number of repeats is 4 and in each repeat the number of convolutional blocks \(M\) is 8.




\subsection{Evaluation metrics}

In this work, in order to evaluate the performance of speech separation results, the models are evaluated on the scale-invariant source-to-noise ratio (SI-SNR), the signal-to-distortion ratio (SDR), signal-to-interference ratio (SIR) and signal-to-artifact ratio (SAR) which are the BBS-eval \cite{vincent2006performance} score, the perceptual evaluation of speech quality (PESQ) \cite{rix2002perceptual} measure and the short-time objective intelligibility (STOI) measure \cite{taal2010short}.

\subsection{Comparison with ideal T-F masks}

In order to compare with the ideal T-F masks, we use the ideal PSM (IPSM), ideal binary mask (IBM) and ideal ratio mask (IRM). These masks are calculated by STFT with 32 ms length hamming window and 16 ms window shift, which is the same as the pre-separation stage. The IPSM is defined in Eq.~\ref{eq4}. The IBM and IRM of source \(s=1,2,...,S\) are defined as following:
\begin{equation}
IBM_s(t,f)=
\begin{cases}
1,\ |X_s(t,f)|>|X_{j\ne s}(t,f)| \\
0,\ otherwise
\end{cases}
\label{eq26}
\end{equation}

\begin{equation}
IRM_s(t,f)=\frac{|X_s(t,f)|}{\sum_{j=1}^S|X_j(t,f)|}
\label{eq27}
\end{equation}

\section{Results}

\subsection{Pre-separation stage}

We firstly evaluate the performance of the pre-separation stage in the T-F domain. Table~\ref{tab:results1} shows the results of SDR, SIR, SAR and PESQ between the uPIT based different speech separation methods on closed (CC) and open (OC) condition. The deep embedding features is denoted by DEF. In Table~\ref{tab:results1}, the "Optimal (Opt.) Assign." means that outputs are optimal assignment. In other words, outputs are with optimal permutation for all of the frames in a utterance. Otherwise, it is the "Default (Def.) Assign.".

\subsubsection{Evaluation of deep embedding features}

From Table~\ref{tab:results1}, we can find that in all objective measures, uPIT+DEF methods all outperform the uPIT method no matter what \(\lambda\) is. These results indicate that the uPIT based separation method with deep embedding features can improve the performance of speaker-independent speech separation. This is because that these deep embedding features are deep representations for the mixture amplitude spectrum, which contain the potential information of each target source so that they can effectively estimate the masks of target sources. Therefore, these deep embedding features are discriminative features for speech separation.

\begin{figure}[t]
	\centering
	\begin{minipage}[t]{0.38\textwidth}
		\includegraphics[width=\linewidth]{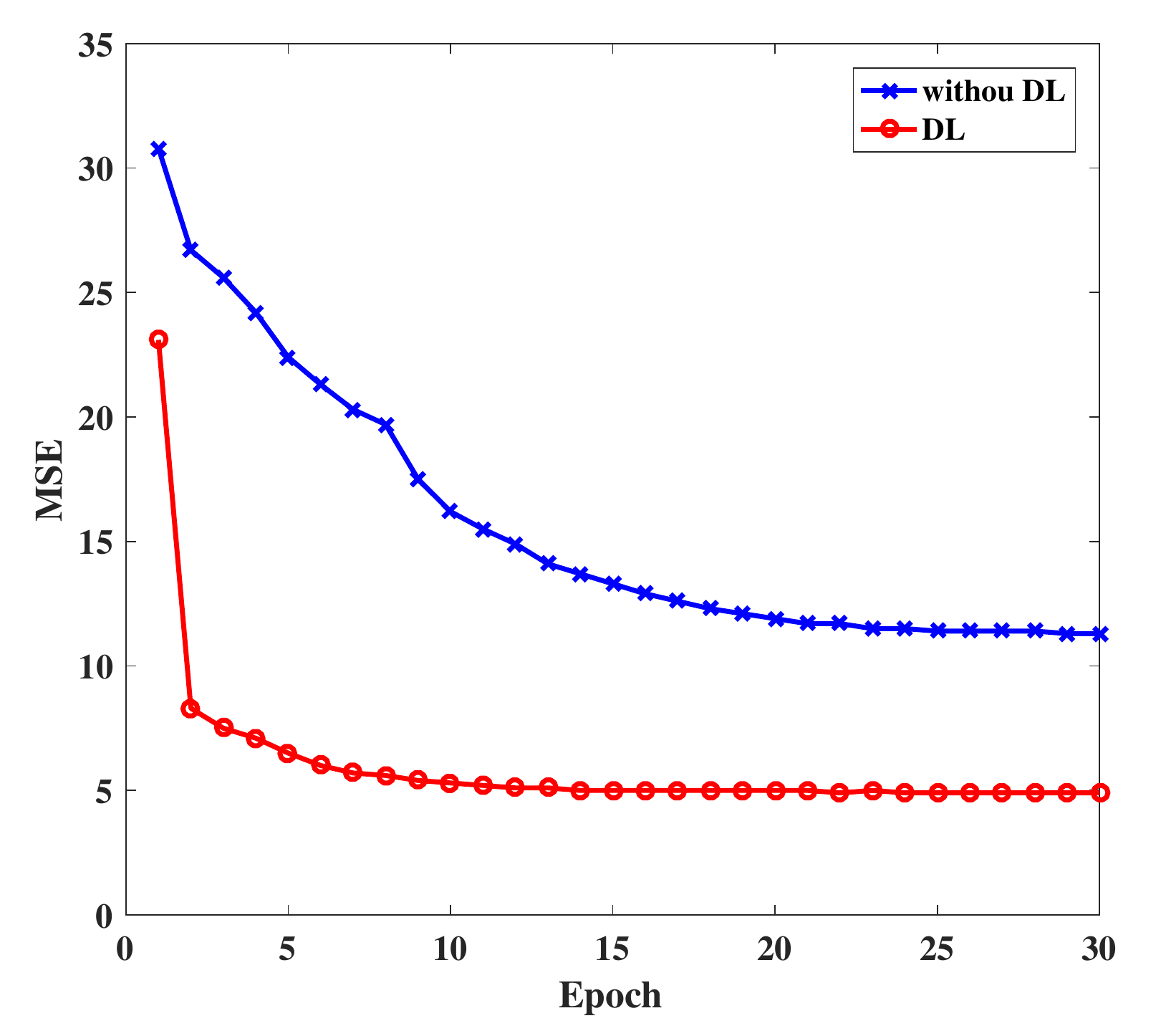}
	\end{minipage}
	\caption{MSE over epochs on the WSJ0-2mix with and without DL training method based on uPIT+DEF.}
	\label{fig:MSE}
\end{figure}

\subsubsection{Evaluation of discriminative learning}

The aim of discriminative learning is to maximize the distance between different sources and minimize the distance between same sources, simultaneously. 

Compared with the uPIT+DEF, uPIT+DEF+DL (discriminative learning is utilized) achieves better performance in the majority of cases, except for the PESQ measure. These results indicate that the discriminative learning can improve the performance of speech separation. Meanwhile, especially for the BSS-eval evaluation metrics (SDR, SIR and SAR), using discriminative learning can gets a better result. The reason is that the discriminative learning increases the dissimilarity between different speakers so that the possibility of remixing the interferences can be reduced. Although the performance of uPIT+DEF+DL is slightly worse than uPIT+DEF for PESQ measure, it is also comparable to the uPIT+DEF and significantly better than the uPIT.

Fig.~\ref{fig:MSE} shows the MSE over epochs on the WSJ0-2mix with and without DL training method based on uPIT+DEF. From Fig.~\ref{fig:MSE} we can find that the DL based separation can be faster convergent than the without DL method. This result indicates the effectiveness of DL.


\setlength{\tabcolsep}{1.5mm}{
	\begin{table}[t]
		\caption{The results of SI-SNR, SDR, PESQ and STOI for the proposed method in time domain and the T-F domain based methods on WSJ0-2mix dataset. They are all in the default assignment and open condition.}
		\label{tab:results2}
		\centering
		\begin{tabular}{c|c|c|c|c}
			\midrule
			Methods              & SI-SNR(dB) & SDR(dB)  & PESQ &STOI(\%)\\ \midrule
			uPIT                & 9.5    & 10.1 & 2.58 & 87.64\\ \midrule
			uPIT+DEF(\(\lambda=0.01\))            & 10.1   & 10.7 & 2.75 & 88.62\\ 
			uPIT+DEF(\(\lambda=0.05\))            & 10.1   & 10.8 & 2.75 & 88.59\\ 
			uPIT+DEF(\(\lambda=0.1\))            & 10.1   & 10.7 & 2.74 & 88.55\\ 
			uPIT+DEF+DL(\(\lambda=0.05\))        & 10.3   & 10.8 & 2.73 & 88.69\\ \midrule
			uPIT+DEF+DL+E2EPF & \multirow{2}*{{16.6}}   & \multirow{2}*{{17.0}}   & \multirow{2}*{{3.41}} & \multirow{2}*{{95.38}}\\
			(proposed) & & & &\\ 
			uPIT+DEF+DL+E2EPF & \multirow{2}*{\textbf{16.9}}   & \multirow{2}*{\textbf{17.3}}   & \multirow{2}*{\textbf{3.43}} & \multirow{2}*{\textbf{95.39}}\\
			+attention (proposed) & & & &\\ \midrule
		\end{tabular}
\end{table}}

\begin{figure*}[t]
	\centering
	\subfigure[SI-SNR]{
		\begin{minipage}[t]{0.45\linewidth}
			\centering
			\includegraphics[width=2.7in]{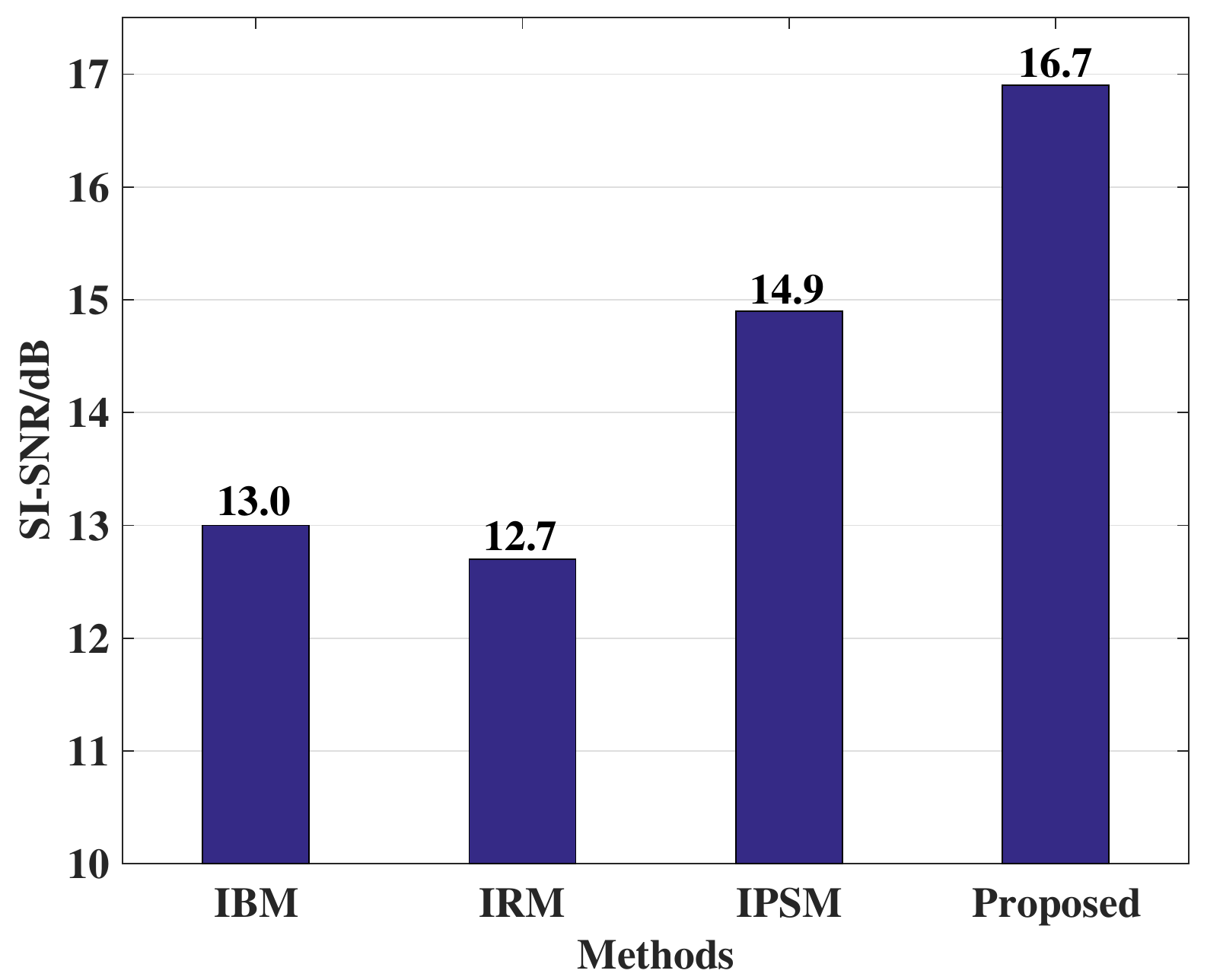}
		\end{minipage}%
	}%
	\subfigure[SDR]{
		\begin{minipage}[t]{0.45\linewidth}
			\centering
			\includegraphics[width=2.7in]{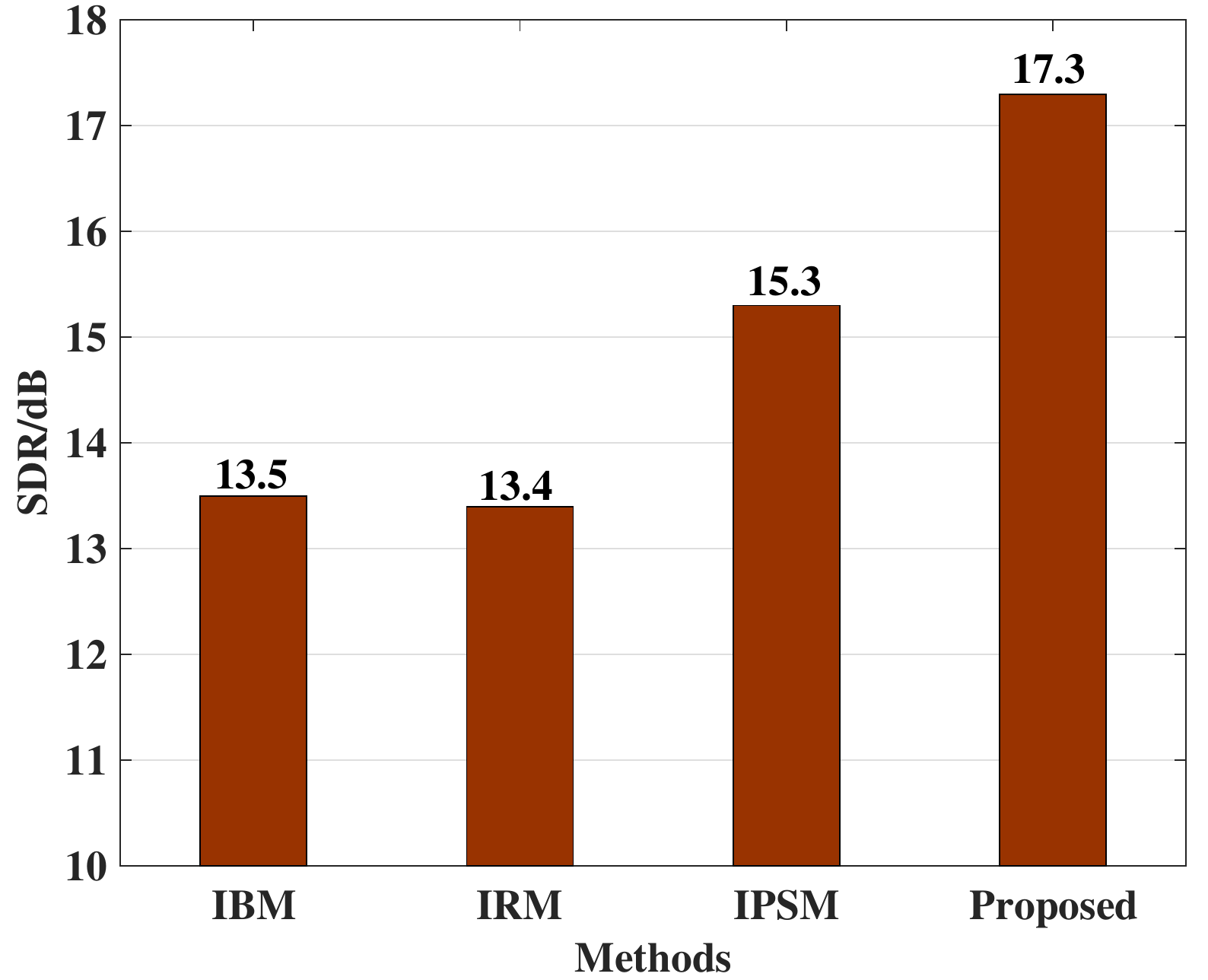}
		\end{minipage}%
	}%
	
	\subfigure[PESQ]{
		\begin{minipage}[t]{0.45\linewidth}
			\centering
			\includegraphics[width=2.7in]{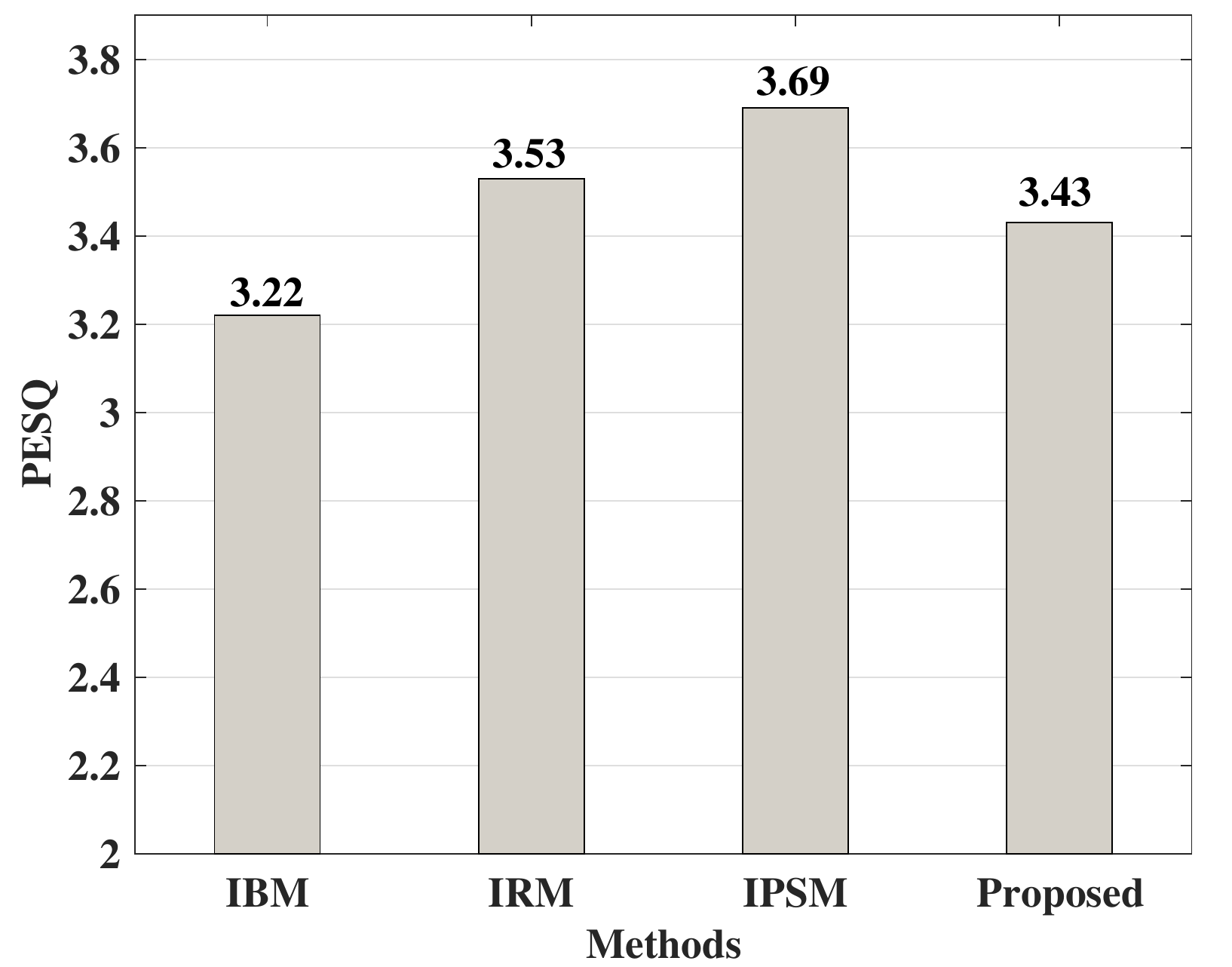}
		\end{minipage}
	}%
	\subfigure[STOI]{
		\begin{minipage}[t]{0.45\linewidth}
			\centering
			\includegraphics[width=2.7in]{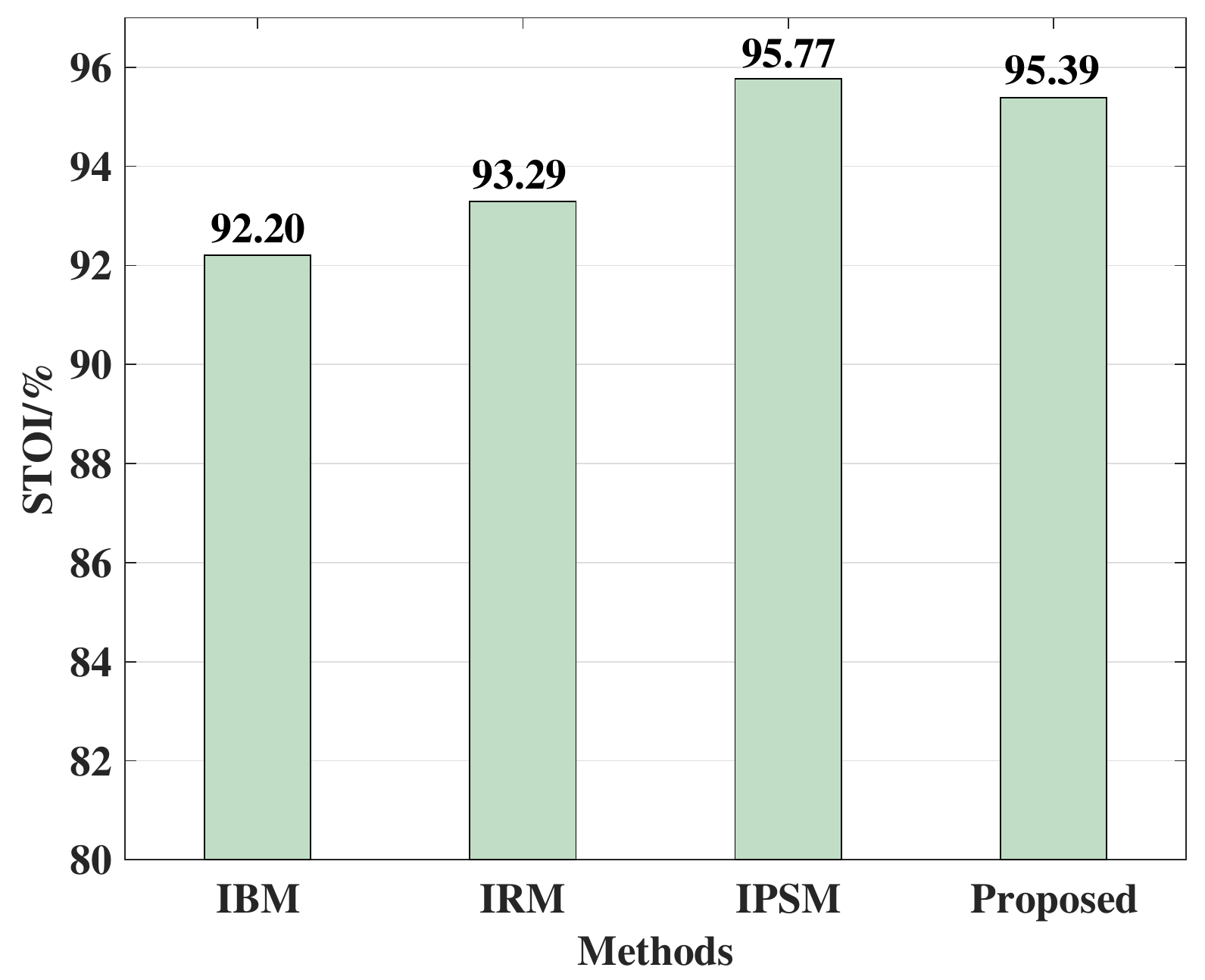}
		\end{minipage}
	}%
	\centering
	\caption{The results of SI-SNR, SDR, PESQ and STOI for the proposed method uPIT+DEF+DL+E2EPF+attention and ideal masks on WSJ0-2mix dataset. (a) The SI-SNR result. (b) The SDR results. (c) The PESQ results. (d) The STOI results.}
	\label{fig:results} 
\end{figure*}



\subsection{Comparison of the proposed end-to-end post-filter method with the uPIT based methods}

Table~\ref{tab:results2} shows the results of SI-SNR, SDR, PESQ and STOI for the proposed method in time domain and the T-F domain uPIT based methods. They are all in the default assignment and open condition. In this study, we extend uPIT+DEF+DL and propose the end-to-end post-filter method for monaural speech separation with deep attention fusion features (uPIT+DEF+DL+E2EPF+attention). 

From the Table~\ref{tab:results2} we can know that when the speech signals separated by the pre-stage (uPIT+DEF+DL method) are processed by the end-to-end post-filter, the performance of speech separation can be improved significantly. More specifically, compared with the uPIT+DEF+DL, our proposed speech separation method uPIT+DEF+DL+E2EPF+attention obtains 6.6 dB increment in SI-SNR, 6.5 dB increment in SDR, 0.7 increment in PESQ and 6.7\% increment in STOI. The reason of the large improvement is that the uPIT+DEF+DL method does the speech separation in the T-F domain and it only enhances the amplitude spectrum, while the phase spectrum is left unchanged. In other words, the uPIT+DEF+DL method utilizes the separated magnitude spectrum and the mixture phase spectrum to reconstruct the each source signals by ISTFT. However, the separated magnitude spectrum and the mixture phase spectrum are mismatched, which damages the performance of speech separation. As for our proposed uPIT+DEF+DL+E2EPF+attention method, the pre-separation stage does the speech separation in the T-F domain to separate the mixture preliminarily. At the end-to-end post-filter stage, in order to improve the performance of speech separation, it applies the waveform as the input features. The waveform contains all of the information of the mixture signals, including magnitude spectrum and phase spectrum. Therefore, this stage enhances the magnitude spectrum and phase spectrum, simultaneously. In addition, to reduce the complexity and size of the end-to-end post-filter model, at the end-to-end post-filter stage, all structures are CNN. 

\begin{table*}[t]
	\caption{The SDR, PESQ and STOI results of different separation methods for different gender combinations on WSJ0-2mix dataset. They are all in the default assignment and open condition.}
	\label{tab:results3}
	\centering
	\begin{tabular}{c|c|c|c|c|c|c|c|c|c}
		\midrule
		\multirow{2}{*}{Methods} & \multicolumn{3}{c|}{Male-Female} & \multicolumn{3}{c|}{Female-Female} & \multicolumn{3}{c}{Male-Male} \\ \cline{2-10} 
		& SDR(dB)          & PESQ     & STOI(\%)        & SDR(dB)           & PESQ     & STOI(\%)        & SDR(dB)         & PESQ     & STOI(\%)       \\ \midrule
		uPIT           & 11.7             & 2.74   & 89.95          & 8.6               & 2.43    &84.59         & 9.9             & 2.37     &85.24       \\ \midrule
		uPIT+DEF(\(\lambda=0.01\))              & 12.3             & 2.92    &90.71         & 9.0               & 2.56   & 85.26          & 8.7             & 2.55     &86.81       \\
		uPIT+DEF(\(\lambda=0.05\))              & 12.5             & 2.91    &90.71         & 9.3               & 2.56   & 85.38          & 8.8             & 2.55     &86.65       \\
		uPIT+DEF(\(\lambda=0.1\))              & 12.3             & 2.91    &90.72         & 9.2               & 2.57   & 85.68          & 8.5             & 2.53     &86.30       \\
		uPIT+DEF+DL(\(\lambda=0.05\))          & 12.5             & 2.90     &90.76        & 9.3               & 2.55    &85.50         & 8.8             & 2.53     &86.81       \\ \midrule
		uPIT+DEF+DL+E2EPF    & \multirow{2}*{19.8}    & \multirow{2}*{3.63}  &\multirow{2}*{97.44}   & \multirow{2}*{13.1}     & \multirow{2}*{\textbf{3.10}}  & \multirow{2}*{\textbf{91.48}}   & \multirow{2}*{14.2}   & \multirow{2}*{3.20}  & \multirow{2}*{\textbf{93.97}} \\ 
		(proposed) &&&&&&&&& \\
		uPIT+DEF+DL+E2EPF    & \multirow{2}*{\textbf{20.2}}    & \multirow{2}*{\textbf{3.65}}  & \multirow{2}*{\textbf{97.50}}   & \multirow{2}*{\textbf{13.3}}     & \multirow{2}*{\textbf{3.10}}  & \multirow{2}*{{91.43}}   & \multirow{2}*{\textbf{14.5}}   & \multirow{2}*{\textbf{3.23}}  & \multirow{2}*{{93.92}} \\ 
		+attention(proposed) &&&&&&&&& \\ \midrule
		IBM &13.7	&3.22&	92.05	&14.1	&3.24	&92.29	&12.7	&3.19	&92.42
		\\
		IRM &13.6	&3.53	&93.05	&14.0	&3.56	&93.89	&12.6	&3.52	&93.35
		\\
		IPSM &15.5	&3.68	&95.74	&15.9	&3.67	&95.70	&14.6	&3.70	&95.87
		\\ \midrule
	\end{tabular}
\end{table*}

\begin{table}[]
	\caption{Comparison with other state-of-the-art systems on WSJ0-2mix dataset.}
	\label{tab:results4}
	\centering
	\begin{tabular}{c|c|c|c|c}
		\midrule
		Methods& SI-SNR(dB)    & SDR(dB)       & PESQ          & STOI(\%)       \\ \midrule
		Mixture           & 0.0           & 0.15          & 2.02          & 74.40          \\ \midrule
		DC\cite{Hershey2016Deep}          & -             & 6.7           & -             & -              \\ 
		DC++\cite{isik2016single}          & 10.8             & -           & -             & -              \\ 
		uPIT\cite{Kolbaek2017Multitalker}        & -             & 10.2          & 2.84          & -              \\ 
		SDC-MLT-Grid\cite{xu2018shifted} & -             & 10.7          & -             & -              \\ 
		CASA-E2E\cite{liu2018casa}    &-               & 11.2              & -              & -               \\
		Chimera++\cite{wang2018alternative}    & -              &11.7               &-               &-                \\
		Wang et al.\cite{wang2019pitch}         &-               &12.0               &-               &-                \\
		TasNet\cite{luo2018arXivtasnet} & 14.6          & 15.2          & 3.25          & -              \\ 
		Conv-TasNet\cite{luo2019conv} & 15.3          & 15.8          & 3.24          & -              \\ 
		Wang et al.\cite{wang2019deep}         &15.3               &15.8               &3.36               &-                \\ \midrule
		uPIT+DEF+DL+E2EPF & \multirow{2}*{\textbf{16.9}}   & \multirow{2}*{\textbf{17.3}}   & \multirow{2}*{\textbf{3.43}} & \multirow{2}*{\textbf{95.39}}\\
		+attention (proposed) & & & &\\ \midrule
		IBM               & 13.0          & 13.5          & 3.22          & 92.20           \\ 
		IRM               & 12.7          & 13.4          & 3.53          & 93.29          \\ 
		IPSM              & 14.9          & 15.3          & 3.69          & 95.77          \\ \midrule
	\end{tabular}
\end{table}

\begin{table}[t]
	\caption{Comparison with other state-of-the-art systems on WSJ0-3mix dataset.}
	\label{tab:results6}
	\centering
	\begin{tabular}{c|c|c|c|c}
		\hline
		Method                                & \(\bigtriangleup\)SI-SNR(dB)   & \(\bigtriangleup\)SDR(dB)      & PESQ          & STOI(\%)       \\ \midrule
		Mixture                               & 0             & 0             & 1.66          & 62.97          \\ \midrule
		uPIT\cite{Kolbaek2017Multitalker}                                  & -             & 7.7           & -             & -              \\ 
		DC++\cite{isik2016single}    & 7.1             & -           & -             & -              \\ 
		DANet\cite{chen2017deep}                                & 8.6           & 8.9           & -          & -              \\ 
		ADANet\cite{luo2018speaker}                                & 9.1           & 9.4           & 2.16          & -              \\ 
		Conv-TasNet\cite{luo2018arXivtasnet}                           & 11.6          & 12.0          & 2.50          & -              \\ \midrule
		uPIT+DEF+DL                           & 7.2           & 8.0           & 2.03          & 74.79          \\ 
		uPIT+DEF+DL+E2EPF & \multirow{2}*{\textbf{12.5}} & \multirow{2}*{\textbf{13.0}} & \multirow{2}*{\textbf{2.70}} & \multirow{2}*{\textbf{87.05}} \\ 
		+attention(proposed) & & & &\\ \midrule
		IBM                                   & 12.0          & 12.3          & 2.80          & 85.82          \\ 
		IRM                                   & 12.4          & 12.8          & 3.50          & 91.92          \\ 
		IPSM                                  & 15.1          & 15.4          & 3.48          & 93.20          \\ \midrule
	\end{tabular}
\end{table}


\subsection{Evaluation of the deep attention fusion features}

In Table~\ref{tab:results2}, Table~\ref{tab:results3} and Table~\ref{tab:results4}, the '\emph{uPIT+DEF+DL+E2EPF}' means without the module of deep attention fusion, the '\emph{uPIT+DEF+DL+E2EPF+attention}' means with the module of deep attention fusion.

From Table~\ref{tab:results2} we can find that when the deep attention fusion features are applied, the performance of speech separation can be improved. More specifically, compare to the uPIT+DEF+DL+E2EPF method, the uPIT+DEF+DL+E2EPF+attention can acquire 0.3 dB increment for both SI-SNR and SDR evaluation metrics. The reason is that these deep attention fusion features are extracted by the attention module, which computes the similarity between the mixture and the pre-separated signals. Therefore, these deep attention fusion features can make the separation model can pay more attention to the pre-separated signals. So they are conducive to help reduce the residual interference and enhance the pre-separated speech so that the performance of speech separation can be improved. These results prove that deep attention fusion features are effective for speech separation.

Examples of separated speech for the baseline and our proposed method are available online\footnote{Available online at https://github.com/fchest/wave-samples}.

\subsection{Comparison of the proposed method with the ideal masks}

In order to make a comparison of our proposed method with the ideal masks, Fig.~\ref{fig:results} shows the results of SI-SNR, SDR, PESQ and STOI for our proposed method and the the ideal masks. 

From Fig.~\ref{fig:results}, several observations can be found. Firstly, IPSM has the best performance compared with the other ideal masks (IBM and IRM) in all evaluation metrics. This is because that the IPSM is a phase sensitive mask, which makes full use of the phase information. Therefore, the phase is very important for speech separation. Secondly, as for SI-SNR and SDR evaluation metrics as shown in Fig.~\ref{fig:results} (a) and (b), our proposed method uPIT+DEF+DL+E2EPF+attention acquires the best performance compared with the ideal masks. If only enhances the magnitude spectrum and leaves the phase spectrum unchanged, these ideal masks are the limitation performance of speech separation. However, the performance of our proposed method is better than these ideal masks, which reveals that our proposed method can separate the mixture very well. Finally, as for PESQ evaluation metric Fig.~\ref{fig:results} (c), although the performance of the proposed method is slightly worse than IPSM, it is still better than IBM and comparable to IRM. And as for the STOI  evaluation metric Fig.~\ref{fig:results} (d), our proposed method is comparable to the IPSM and outperforms the IBM and IRM. Therefore, these results indicate the effectiveness of our proposed method for speech separation.

\subsection{Comparison with different gender combinations}

Table~\ref{tab:results3} compares the results of uPIT based speech separation methods for different gender combinations. Male-female combinations can acquire a better performance than female-female and male-male combinations for all of speech separation methods in Table~\ref{tab:results3}. This is because that compared with the same gender combinations, different gender combinations have larger differences for speech features, for example pitch. Therefore, the same gender combinations speech is more difficult to separate. However, our proposed method can achieve better results than other methods for all of the gender combinations, especially for the same gender combinations. These results indicate that our proposed method is effective for speech separation.

%

\subsection{Comparison with other state-of-the-art methods}

In order to compare the separation results of our proposed method with previous methods, Table~\ref{tab:results4} shows the performance of our proposed method uPIT+DEF+DL+E2EPF+attention and other state-of-the-art methods on the same WSJ0-2mix dataset. For all methods, the best reported results are listed and they are all in the default assignment and open condition. Note that, for \cite{Kolbaek2017Multitalker,Hershey2016Deep,xu2018shifted,liu2018casa,wang2018alternative,luo2018arXivtasnet,luo2019conv,wang2019deep} methods are use SDR improvements results. To compare equally, their final results are add 0.2 dB although the SDR result of the mixture is only about 0.15 dB. In this table, the missing values are because they are unreported in their corresponding study.

As for DC\cite{Hershey2016Deep}, DC++\cite{isik2016single}, uPIT\cite{Kolbaek2017Multitalker}, SDC--MLT-Grid\cite{xu2018shifted} and CASA-E2E\cite{liu2018casa}, they all do the speech separation in the T-F domain with no phase enhancement. Their performance are slightly worse than the other speech separation methods. TasNet\cite{luo2018arXivtasnet} and Conv-TasNet\cite{luo2019conv} extend uPIT to the time domain and use the TCN for separation, which acquire quite good results. Note that the TasNet\cite{luo2018arXivtasnet} does not use the  prior knowledge of the pre-separated speech. From Table~\ref{tab:results4} we can find that our proposed method acquires the best performance, which indicate the effectiveness of our proposed method. The reason is that our proposed method can make full use of the prior knowledge of the pre-separated speech to help reduce the residual interference. In order to address the mismatch problem of magnitude and phase, our proposed E2EPF utilizes the waveform as the input feature, which can enhance the magnitude and phase simultaneously. In addition, the deep attention fusion features are applied to E2EPF so that the E2EPF can pay more attention to the pre-separated speech. Therefore, the E2EPF can enhance the separated speech very well and the performance of speech separation can be improved.

Table~\ref{tab:results6} shows the results of our proposed method uPIT+DEF+DL+E2EPF+attention and other state-of-the-art methods on the same WSJ0-3mix dataset. As the SI-SNR and SDR of the mixture in WSJ0-3mix dataset are negatives, we use the \(\bigtriangleup\)SI-SNR and \(\bigtriangleup\)SDR as the evaluation metrics. From Table~\ref{tab:results6} we can find that our proposed method uPIT+DEF+DL+E2EPF+attention outperforms other separation systems on the WSJ0-3mix dataset. These results indicate that our proposed method is effective for speech separation.

\section{Discussions}

The above experimental results show that our proposed end-to-end post-filter method with deep attention fusion features is effective for speaker independent speech separation. We can make some interesting observations as follows.

Our proposed end-to-end post-filter method can further reduce the residual interference and improve the performance of speech separation. The performance of the pre-separation stage uPIT+DEF+DL method outperforms the uPIT method but it still needs to be improved. This is because that the separated speech by this stage may still contain the residual interference. In addition, it uses the mismatched mixture phase and the enhanced magnitude to reconstruct the separated speech, which damages the separation performance. When the proposed end-to-end post-filter method is utilized, the separation performance can be improved. The reason is that the end-to-end post-filter makes full use of the prior knowledge of pre-separated speech so that it can reduce the residual interference and improve the separation performance. Besides, it utilizes the waveform as the input features, which includes the magnitude and phase. Therefore, when it enhances the waveform, the amplitude and phase can be enhanced simultaneously. So our proposed method can address the mismatch problem of the magnitude and phase.

The deep attention fusion features are conducive to speech separation. Compared to the uPIT+DEF+DL+E2EPF method (without deep attention fusion features), the proposed uPIT+DEF+DL+E2EPF+attention can acquire a better speech separation result. The reason is that these deep attention fusion features are extracted by an attention module that computes the similarity between the mixture and pre-separated signals. Therefore, the end-to-end post-filter can pay more attention to the pre-separated signals so that the residual interference can be reduced and the pre-separated speech can be enhanced further. 

In summary, our proposed end-to-end post-filter method can further reduce the residual interference. Furthermore, the deep attention fusion features are applied to improve the performance of speech separation.


\section{Conclusion}


In this paper, we presented an end-to-end post-filter method for monaural speech separation, which utilized the deep attention fusion features. The uPIT+DEF+DL method was applied to separate the mixture speech preliminarily. In order to further reduce the interference, the end-to-end post-filter with the deep attention fusion features was proposed. Our experiments were conducted on WSJ0-2mix and WSJ0-3mix dataset. Results showed that the proposed method was effective for speaker-independent speech separation. In the future, we will extend the proposed method to multi-channel speech separation, which could use the spatial information to improve the performance of speech separation.

\ifCLASSOPTIONcaptionsoff
  \newpage
\fi



%

%
%

\bibliography{references}{}
\bibliographystyle{IEEEtran}

%

\begin{IEEEbiography}[{\includegraphics[width=1.1in,height=1.25in,clip,keepaspectratio]{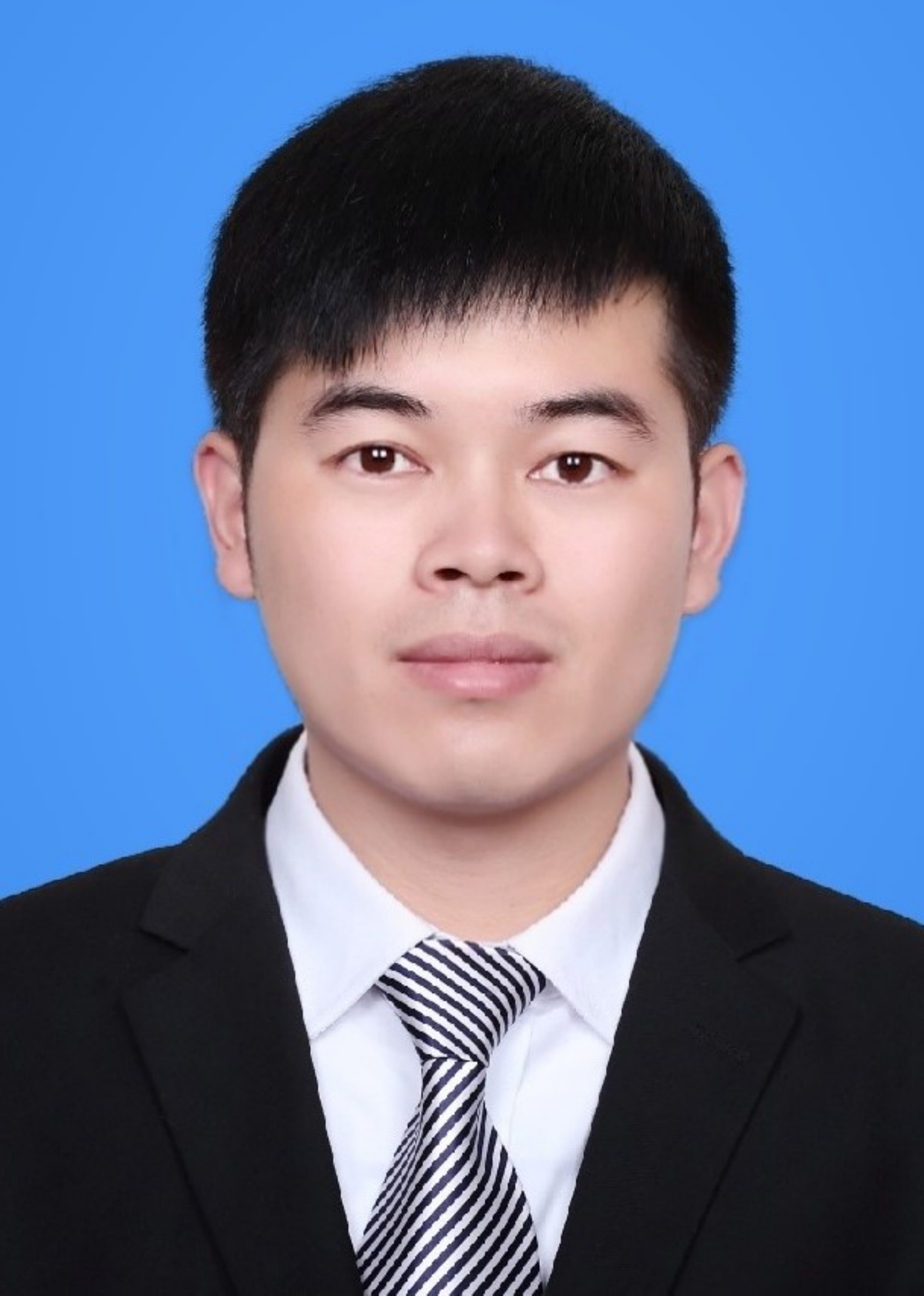}}]{Cunhang Fan}
received the B.S. degree from the Beijing University of Chemical Technology (BUCT), Beijing, China, in 2016. He is currently working toward the Ph.D degree with the National Laboratory of Pattern Recognition (NLPR), Institute of Automation, Chinese Academy of Sciences (CASIA), Beijing, China. His current research interests include speech separation, speech enhancement, speech recognition and speech signal processing.
\end{IEEEbiography}
\begin{IEEEbiography}[{\includegraphics[width=1.1in,height=1.25in,clip,keepaspectratio]{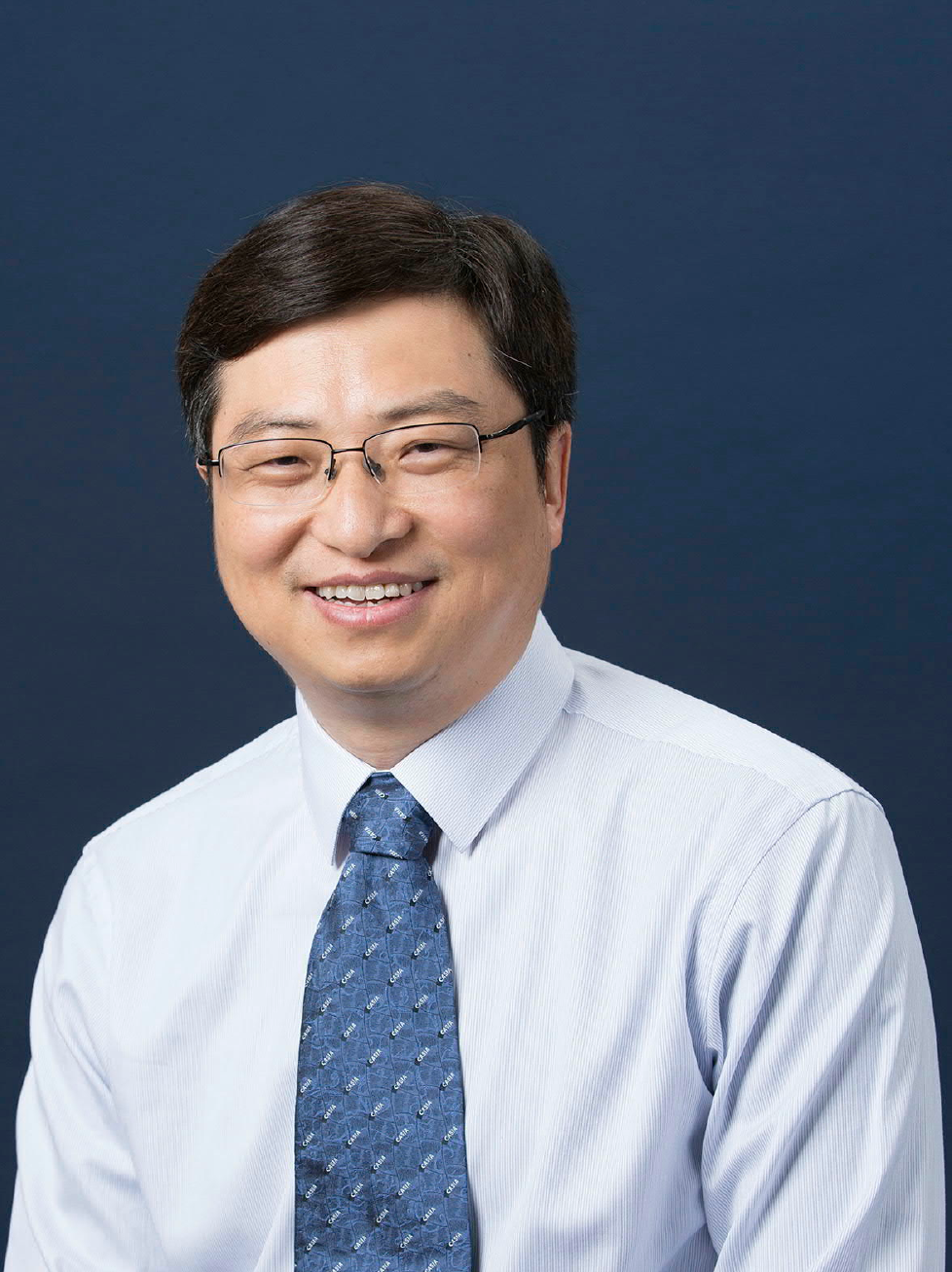}}]{Jianhua Tao}
	received the Ph.D. degree from Tsinghua University, Beijing, China, in 2001, and the M.S. degree from Nanjing University, Nanjing, China, in 1996. He is currently a Professor with
	NLPR, Institute of Automation, Chinese Academy of Sciences, Beijing, China. He has authored or coauthored more than eighty papers on major journals and proceedings including the IEEE TRANSACTIONS ON AUDIO, SPEECH, AND LANGUAGE PROCESSING. His current research interests include speech recognition, speech synthesis and coding methods, human–computer interaction, multimedia information processing, and pattern recognition. He is the Chair or Program Committee Member for several major conferences,
	including ICPR, ACII, ICMI, ISCSLP, NCMMSC, etc. He is also the Steering Committee Member for the IEEE TRANSACTIONS ON AFFECTIVE COMPUTING, an Associate Editor for Journal on Multimodal User Interface and International Journal on Synthetic Emotions, and the Deputy Editor-in-Chief for Chinese Journal of Phonetics. He was the recipient of several awards from the important conferences, such as Eurospeech, NCMMSC, etc.
\end{IEEEbiography}
\begin{IEEEbiography}[{\includegraphics[width=1.1in,height=1.25in,clip,keepaspectratio]{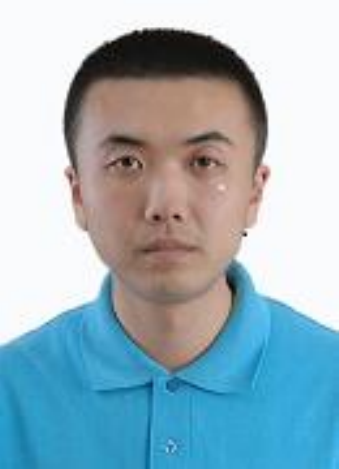}}]{Bin Liu}
	received his the B.S. degree and the M.S. degree from Beijing institute of technology (BIT), Beijing, in 2007 and 2009 respectively. He received Ph.D. degree from the National Laboratory of Pattern Recognition, Institute of Automation, Chinese Academy of Sciences, Beijing, in 2015. He is currently an Associate Professor in the National Laboratory of Pattern Recognition, Institute of Automation, Chinese Academy of Sciences, Beijing. His current research interests include affective computing and audio signal processing.
\end{IEEEbiography}
\begin{IEEEbiography}[{\includegraphics[width=1.1in,height=1.25in,clip,keepaspectratio]{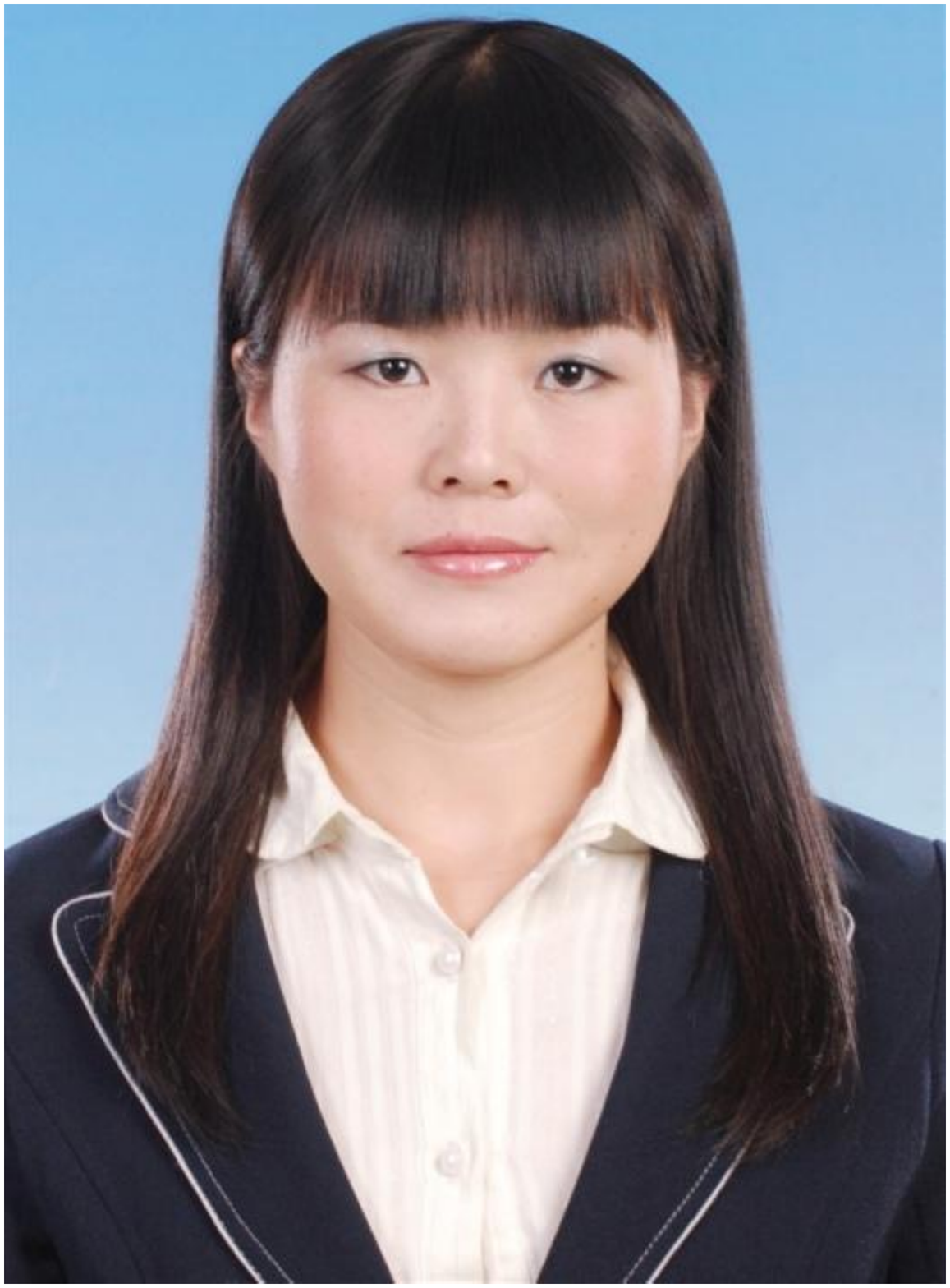}}]{Jiangyan Yi}
	received the Ph.D. degree from the University of Chinese Academy of Sciences, Beijing, China, in 2018, and the M.A. degree fromthe Graduate School of Chinese Academy of Social Sciences, Beijing, China, in 2010. She was a Senior R\&D Engineer with Alibaba Group during 2011 to 2014. She is currently an Assistant Professor with the National Laboratory of Pattern Recognition, Institute of Automation, Chinese Academy of Sciences, Beijing, China. Her current research interests include speech processing, speech recognition, distributed computing, deep learning, and transfer learning. 
\end{IEEEbiography}
\begin{IEEEbiography}[{\includegraphics[width=1.1in,height=1.25in,clip,keepaspectratio]{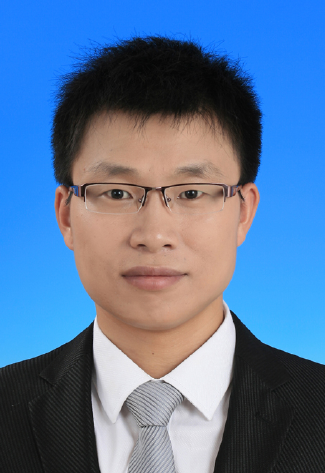}}]{Zhengqi Wen}
	received the B.S. degree from the University of Science and Technology of China, Hefei, China, in 2008, and the Ph.D. degree from the Chinese Academy of Sciences, Beijing, China, in 2013. He is currently an Associate Professor with the National Laboratory of Pattern Recognition, Institute of Automation, Chinese Academy of Sciences. His current research interests include speech processing, speech recognition, and speech synthesis.
\end{IEEEbiography}
\begin{IEEEbiography}[{\includegraphics[width=1.1in,height=1.25in,clip,keepaspectratio]{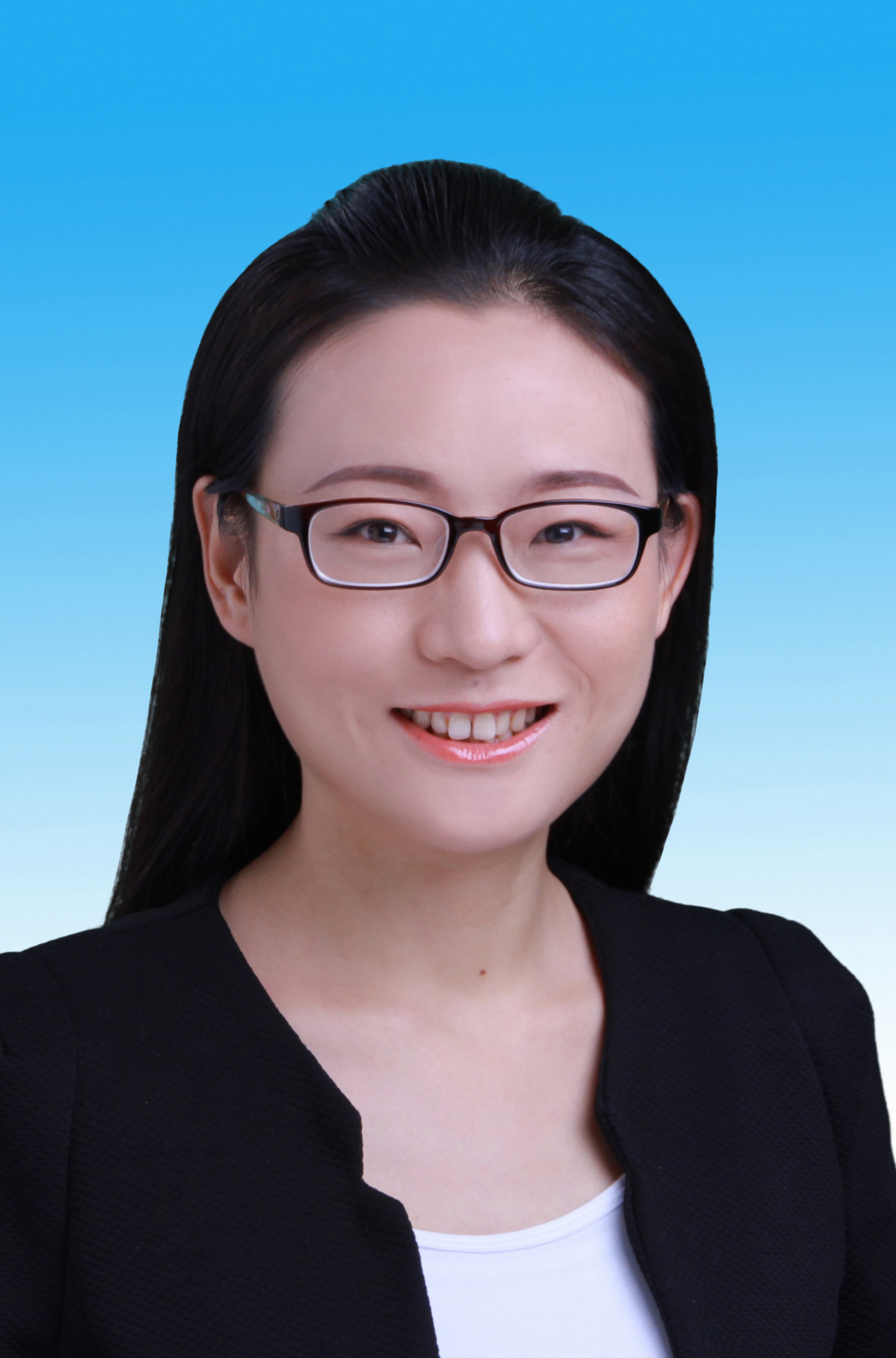}}]{Xuefei Liu}
	 received the Ph.D. degree from the Graduate School of Chinese Academy of Social Sciences, Beijing, China, in 2016, and the M.A. degree from Beijing Normal University, Beijing, China, in 2013. She is currently an Assistant Professor with the National Laboratory of Pattern Recognition, Institute of Automation, Chinese Academy of Sciences, Beijing, China. Her current research interests include Corpus Construction and experimental phonetics. 
\end{IEEEbiography}




\end{document}